\documentclass[preprintnumbers,prd,twocolumn,tightenlines,floatfix,showpacs,amssymb,nofootinbib,superscriptaddress,a4paper]{revtex4}

\usepackage{pictex}
\usepackage[dvips]{graphicx}
\usepackage{amsmath}

\usepackage{amsmath}
\usepackage{amsfonts}
\usepackage{amsthm}

\newcommand{\del}{\partial}
\DeclareMathOperator{\Tr}{Tr}
\DeclareMathOperator{\re}{Re}
\newcommand{\Dslash}{\!\!\!\!/\,}
\newcommand{\pslash}{\!\!\!/}
\newcommand{\nn}{\nonumber}
\newcommand{\half}{\frac{1}{2}}
\newcommand{\quarter}{\frac{1}{4}}
\newcommand{\eqn}[1]{ \begin{equation} #1 \end{equation} }
\newcommand{\expectation}[1]{\langle #1 \rangle}

\newcommand{\link}{
\setlength{\unitlength}{14pt}
\begin{picture}(1,1)(0,0)
\linethickness{0.25pt}
\put(0,0){\circle*{0.15}}
\put(0,0){\vector(1,0){1}}
\put(1,0){\circle{0.14}}
\end{picture}}
\newcommand{\stapleup}{
\setlength{\unitlength}{14pt}
\begin{picture}(1,1)(0,0)
\linethickness{0.25pt}
\put(0,0){\circle*{0.15}}
\put(0,0){\vector(0,1){1}}
\put(0,1){\vector(1,0){1}}
\put(1,1){\vector(0,-1){1}}
\put(1,0){\circle{0.14}}
\end{picture}}
\newcommand{\stapledown}{
\raisebox{-14pt}{
\setlength{\unitlength}{14pt}
\begin{picture}(1,1)(0,-1)
\linethickness{0.25pt}
\put(0,0){\circle*{0.15}}
\put(0,0){\vector(0,-1){1}}
\put(0,-1){\vector(1,0){1}}
\put(1,-1){\vector(0,1){1}}
\put(1,0){\circle{0.14}}
\end{picture}}}

\begin{document}
\preprint{ADP-07-06/T646}

\title{Unquenching effects in the quark and gluon propagator}

\author{Waseem Kamleh}
\affiliation{Special Research Centre for the Subatomic Structure of Matter and Department of Physics, University of Adelaide 5005, Australia. }
\affiliation{School of Mathematics, Trinity College, Dublin 2, Ireland. }
\author{Patrick O. Bowman}
\affiliation{Centre of Theoretical Chemistry and Physics, 
Institute of Fundamental Sciences, Massey University (Auckland), 
Private Bag 102904, NSMSC, Auckland NZ}
\author{Derek B. Leinweber}
\author{Anthony G. Williams}
\affiliation{Special Research Centre for the Subatomic Structure of Matter and Department of Physics, University of Adelaide 5005, Australia. }
\author{Jianbo Zhang}
\affiliation{Special Research Centre for the Subatomic Structure of Matter and Department of Physics, University of Adelaide 5005, Australia. }
\affiliation{ZIMP and Department of Physics, Zhejiang University, Hangzhou 310027, China.}

\begin{abstract}
In this work we examine the Fat-Link Irrelevant Clover (FLIC) overlap
quark propagator and the gluon propagator on both dynamical and
quenched lattices.  The tadpole-improved Luscher-Weisz gauge action is
used in both cases. The dynamical gauge fields use the FLIC fermion
action for the sea quark contribution.  We observe that the presence of
sea quarks causes a suppression of the mass function, quark
renormalization function and gluon dressing function in the
infrared. The ultraviolet physics is unaffected.

\end{abstract}


\maketitle

Quark and gluon propagators are fundamental quantities in QCD encoding
the rich nonperturbative and perturbative properties of QCD. In
previous studies, the Fat-Link Irrelevant Clover (FLIC) overlap quark
propagator and the gluon propagator were examined on quenched
lattices\cite{kamleh-overlap,gluon1,gluon2}. In this work we study the
effects of two flavor dynamical FLIC fermions~\cite{Kamleh:2004xk} on these
quantities for the first time.

The use of smeared or ``fat'' links in lattice fermion actions has
been of interest for some time \cite{Degrand} and FLIC fermions
\cite{zanotti-hadron,kamleh-spin} have shown a number of promising
advantages over standard lattice actions.  The FLIC fermion action is
a Wilson-clover type fermion action in which the irrelevant operators
of the Wilson and clover terms are constructed using fat links, while
the relevant operators use the untouched (thin) gauge links.  By
smearing only the irrelevant, higher dimensional terms in the action,
and leaving the relevant dimension-four operators untouched, short
distance quark and gluon interactions are retained.

Fat links may be created by any number of smearing prescriptions
including APE smearing \cite{ape-one,ape-two,derek-smooth,ape-MIT},
hypercubic smearing \cite{hasenfratz-hyp}, stout-link smearing
\cite{Morningstar:2003gk} or the unit-circle smearing
\cite{Kamleh:2004xk} used herein.

The use of fat links in the FLIC action minimizes the effect of
renormalization on the action improvement terms.  Scaling studies
indicate FLIC fermions provide a new form of nonperturbative
${\mathcal O}(a)$ improvement \cite{Zanotti:2004dr} where
near-continuum results are obtained at finite lattice spacing.

Access to the light quark mass regime is enabled by the improved
chiral properties of the lattice fermion action
\cite{Boinepalli:2004fz}.  In particular, the histogram of the
additive mass renormalization encountered in chiral-symmetry breaking
Wilson-type fermion actions is seen to narrow upon introducing
fat-links in the irrelevant operators.  These benefits facilitate the
generation of the dynamical fermion configurations examined herein.

\section{Continuum Propagators}

We begin by reviewing the formulation of the quark and gluon
propagator on the lattice.
In the continuum, the tree-level $(A_\mu(x) = 0)$ quark propagator is
identified with the (Euclidean space) fermionic Greens function, \eqn{
(\del\pslash + m^0)\Delta_{\rm f}(x,y) = \delta^4(x-y),} where $m^0$
is the bare quark mass. In momentum space this equation is solved
straightforwardly, \eqn{ \tilde{\Delta}_{\rm f}(p) =
\frac{1}{ip\pslash + m^0}. } Denote $S^{(0)}(p) \equiv
\tilde{\Delta}_{\rm f}(p)$ to be the tree-level propagator in momentum
space. Then in the presence of gauge field interactions, define
$S_{\rm bare}(p)$ to be the fourier transform of the (interacting)
fermionic Green's function, \eqn{ (D\Dslash + m^0)\Delta^{\rm
bare}_{\rm f}(x,y) = \delta^4(x-y). } We define the mass function
$M(p)$ and the bare renormalization function $Z(p)$ such that the bare
quark propagator has the form \eqn{ S_{\rm bare}(p) =
\frac{Z(p)}{ip\pslash + M(p)}. } Then for the renormalization point
$\zeta,$ the renormalised quark propagator is given by \eqn{
S_\zeta(p) = \frac{Z_\zeta(p)}{ip\pslash + M(p)} = Z_2(\zeta,a)S_{\rm
bare}(p), } where $Z_\zeta(p)$ is the ($\zeta$-dependent)
renormalization function, and $Z_2(\zeta,a)$ is the wave function
renormalization constant, which depends on $\zeta$ and the regulator
parameter $a.$ The $a$-dependence of $S_{\rm bare}$ is
implicit. $Z_2(\zeta,a)$ is chosen such that \eqn{
Z_\zeta(p)|_{p^2=\zeta^2} = 1. } As $S_\zeta(p)$ is multiplicatively
renormalisable, all of the $\zeta$-dependence is contained within
$Z_\zeta(p),$ that is, the mass function is $\zeta$-independent.

The continuum tree-level gluon propagator in Landau gauge is associated with the following Greens function,
\eqn{ (\delta^{\mu\nu}\del^2 - \del^\mu\del^\nu)\delta_{ab} D^{(0)}{}_{\mu\nu}^{ab}(x,y) = \delta^4(x-y).}
In momentum space the tree-level gluon propagator takes the form,
\eqn{ D^{(0)}{}_{\mu\nu}^{ab}(q) = (\delta_{\mu\nu} - \frac{q_\mu q_\nu}{q^2})\delta^{ab}\frac{1}{q^2}.}
The non-perturbative gluon propagator is defined as the following two-point function,
\eqn{ D_{\mu\nu}^{ab}(x,y) = \expectation{A_\mu^a(x)A_\nu^b(y)}.}
The scalar propagator $D(q^2)$ is related to the full propagator (in momentum space) by
\eqn{ D_{\mu\nu}^{ab}(q) = (\delta_{\mu\nu} - \frac{q_\mu q_\nu}{q^2})\delta^{ab}D(q^2).}
The renormalised scalar propagator $D_\xi(q^2)$ is chosen such that at some momentum scale $\xi,$
\eqn{ q^2 D_\xi(q^2)|_{q^2 = \xi^2} = 1. }
We refer to $q^2 D_\xi(q^2)$ as the gluon dressing function.

\section{FLIC Overlap Fermions}

The overlap formalism~\cite{overlap1,overlap2,overlap3,overlap4} in the vector-like case leads to the following definition of the massless overlap-Dirac operator~\cite{neuberger-massless},
\eqn{ D_{\rm o} = \frac{1}{2a}(1 + \gamma_5 \epsilon(H)).}
Here, $\epsilon(H)$ is the matrix sign function applied to the overlap kernel $H.$ The kernel can be any Hermitian version of the Dirac operator which represents a single fermion species of large negative mass. The standard choice is the Hermitian Wilson-Dirac operator (setting $a=1$),
\eqn{ H_{\rm w} = \gamma_5(\nabla\Dslash + \frac{1}{2}\Delta - m_{\rm w}),}
where $\nabla\Dslash$ is the central covariant finite difference operator, and $\Delta$ is the lattice Laplacian, or Wilson term.

 In this work, as in previous studies, we use the FLIC action~\cite{zanotti-hadron,kamleh-spin} as the overlap kernel\cite{kamleh-overlap}. The FLIC fermion action is a variant of the clover action where the irrelevant operators are constructed using APE-smeared links~\cite{ape-one,ape-two,derek-smooth,ape-MIT}, and mean field improvement~\cite{lepage-mfi} is performed. The Hermitian FLIC operator is given by
\eqn{H_{\rm flic} = \gamma_5(\nabla\Dslash_{\rm mfi} + \frac{1}{2}(\Delta^{\rm fl}_{\rm mfi} - \frac{1}{2}\sigma\cdot F^{\rm fl}_{\rm mfi}) - m_{\rm w}),}
where the presence of fat (smeared) links and/or mean field improvement has been indicated by the super- and subscripts. We choose $\sigma_{\mu\nu}=\frac{i}{2}[\gamma_\mu,\gamma_\nu]$ and use a standard one-loop $F_{\mu\nu},$
\begin{align}
F_{\mu\nu}(x) &= -\frac{i}{2}(C_{\mu\nu}(x) - C^\dagger_{\mu\nu}(x)), \\
\nn C_{\mu\nu}(x) &= \quarter(U_{\mu,\nu}(x) + U_{-\nu,\mu}(x) \\
&\qquad + U_{\nu,-\mu}(x) + U_{-\mu,-\nu}(x)),
\end{align}
where $U_{\mu,\nu}(x)$ is the elementary plaquette in the $+\mu,+\nu$ direction. 

The APE smeared links $U^{\rm fl}_\mu(x)$ constructed from $U_\mu(x)$ by performing $n$ smearing sweeps, where in each sweep we first perform an APE blocking step,
\begin{equation}
V^{(j)}_\mu(x) = (1-\alpha)\ \link + \frac{\alpha}{6} \sum_{\nu \ne \mu}\ \stapleup\ + \stapledown\ , 
\end{equation}
followed by a projection back into $SU(3), U^{(j)}_\mu(x) = {\mathcal P}(V^{(j)}_\mu(x)).$ In this work, the projection is defined by first performing a projection into $U(3),$
\begin{equation}
U'(V) = V [V^\dagger V]^{-\half},
\end{equation}
and then projection into $SU(3),$
\begin{equation}
{\mathcal P}(V) = \frac{1}{\sqrt[3]{\det U'(V)}} U'(V).
\end{equation}
As it is only the product $n\alpha$ that matters~\cite{derek-apesmearing}, we fix $\alpha=0.7$ and only vary $n.$

Mean field improvement is performed by making the replacements
\eqn{ U_\mu(x) \to \frac{U_\mu(x)}{u_0},\quad U^{\rm fl}_\mu(x) \to \frac{U^{\rm fl}_\mu(x)}{u^{\rm fl}_0}, }
where $u_0$ and $u_0^{\rm fl}$ are the mean links for the standard and smeared gauge fields. We calculate the mean link via the fourth root of the average plaquette, 
\eqn{ u_0 = \langle {\rm \frac{1}{3}ReTr } U_{\mu\nu}(x) \rangle_{x,\mu<\nu}^{\frac{1}{4}}. }

\section{Lattice Quark Propagator}

It is easily seen that the continuum massless quark propagator anti-commutes with $\gamma_5,$
\eqn{ \{ \gamma_5, S^{\rm c}_{\rm bare}(p)\big|_{m^0=0} \} = 0. }
A straightforward consequence of the Ginsparg-Wilson relation is the inverse of the overlap Dirac operator satisfies
\eqn{ \{ \gamma_5, D_{\rm o}^{-1} \} = 2\gamma_5. }
Noting that the mean link is a function of $a$ and that $u_0(a), u^{\rm fl}_0(a) \to 1$ as $a \to 0,$ we obtain
\eqn{\lim_{a\to 0} D_{\rm o} = \frac{1}{2 m_{\rm w}} D\Dslash.}
It is then natural to define the (external) massless bare overlap propagator on the lattice as~\cite{overlap4,edwards-study}
\eqn{ S_{\rm bare}(p)|_{m^0=0} \equiv \frac{1}{2 m_{\rm w}} (D_{\rm o}^{-1} - 1), }
as we then have that 
\eqn{ \{ \gamma_5, S_{\rm bare}(p)\big|_{m^0=0} \} = 0, }
as in the continuum case. 
The massive overlap Dirac operator is given by~\cite{neuberger-almostmassless}
\eqn{ D_{\rm o}(\mu) = (1-\mu)D_{\rm o} + \mu,}
with $|\mu| < 1$ representing fermions of mass $\propto \frac{\mu}{1 - \mu}.$ The massive (external) bare overlap propagator is defined as~\cite{edwards-study}
\eqn{ S_{\rm bare}(p) \equiv \frac{1}{2 m_{\rm w}(1-\mu)}(D_{\rm o}^{-1}(\mu)-1), }
and with the identification 
\eqn{\label{eq:baremass}\mu = \frac{m^0}{2 m_{\rm w}} }
satisfies
\eqn{ S^{-1}_{\rm bare}(p) = S^{-1}_{\rm bare}(p)\big|_{m^0=0} + m^0. }

In order to construct $M(p)$ and $Z(p)$ on the lattice, we first define ${\cal B}(p),{\cal C}_\mu(p)$ such that
\eqn{ S_{\rm bare}(p) = -i{\cal C}\Dslash(p) + {\cal B}(p).}
Then 
\begin{align}
{\cal{C}}_{\mu}(p)&= \frac{i}{n_{\rm s}n_{\rm c}}\Tr[\gamma_\mu{S_{\rm bare}(p)}], \\
{\cal{B}}(p)&= \frac{1}{n_{\rm s}n_{\rm c}}\Tr[S^{\rm bare}(p)],
\end{align}
where the trace is over color and spinor indices only, and $n_{\rm s}, n_{\rm c}$ specify the dimension of the spinor and color vector spaces. Now, define the functions $B(p),C_\mu(p)$ such that the inverse of the bare (lattice) quark propagator has the form
\eqn{ S_{\rm bare}^{-1}(p) = iC\Dslash(p) + B(p). }
Then it is easily seen that
\begin{align}
C_\mu &= \frac{{\cal C}_\mu}{{\cal C}^2 + {\cal B}^2}, & B &= \frac{{\cal B}}{{\cal C}^2 + {\cal B}^2},
\end{align}
where ${\cal C}^2 = {\cal C}\cdot{\cal C}.$ 

The kinematical lattice momentum $q_\mu$ is defined such that at tree-level 
\eqn{ (S^{(0)})^{-1}(p) = i q\pslash + m^0,}
that is $q_\mu(p) = C^{(0)}_\mu(p), m^0 = B^{(0)}(p).$ We note that the simple form of these relations is one of the advantages of overlap fermions, as the absence of additive mass renormalization prevents the need for having to perform any tree-level correction~\cite{overlgp} (necessary most other fermions actions~\cite{jon1,jon2,bowman01}), outside of identifying the correct momentum variable $q.$ Now, we define $A(p)$ such that
\eqn{ S_{\rm bare}^{-1}(p) = iq\pslash A(p) + B(p).}
The mass function $M(p)$ and renormalization function $Z(p)$ may then be straightforwardly constructed,
\begin{align}
Z(p) &= \frac{1}{A(p)},& M(p) &= \frac{B(p)}{A(p)}.
\end{align}

\section{Lattice Gluon Propagator}

We use a tadpole-improved plaquette plus rectangle gluon action,
\begin{multline}
 S^{\rm imp}_{\rm gauge} =  \frac{5}{9} \beta \sum_{x \in {\rm L}} \sum_{\mu < \nu} \re \Tr [ (1 - U_{\mu\nu}(x)) - \\ 
 \frac{1}{20 u_0^2}(2 - R^{(2\times 1)}_{\mu\nu}(x) - R^{(1\times 2)}_{\mu\nu}(x))].
\end{multline}
The lattice gauge field may be related to a continuum gauge field through a path ordered exponential,
\eqn{ U_\mu(x) =  {\cal P}e^{ig\int_x^{x+e_\mu} dx_\mu A_\mu(x)}. }
We can recover the lattice gluon field through the following ``mid-point'' definition,
\begin{multline}
A_\mu(x + \half e_\mu) = \frac{1}{2ig}(U_\mu(x) - U^\dagger_\mu(x)) - \\ \frac{1}{6ig}\Tr(U_\mu(x) - U^\dagger_\mu(x)) + {\cal O}(a^2).
\end{multline}

As in the continuum, the lattice gluon propagator in coordinate space is then given by
\eqn{ D_{\mu\nu}^{ab}(x,y) = \expectation{A_\mu^a(x)A_\nu^b(y)}. }
A Fourier transform takes us to momentum space, and we calculate the scalar propagator $D(q^2)$ directly. Tree level improvement of the scalar gluon propagator is performed by identifying the appropriate kinematical momenta $q$ such that
\eqn{ D^{(0)}(q^2) = \frac{1}{q^2}. }
We denote the two different (gluonic and fermionic) kinematical momentum by $q,$ and use context to distinguish them.

\begin{table}[b]
\centering
\begin{ruledtabular}
\begin{tabular}{ccccccc}
Volume &$\beta$ & $\kappa_{\rm sea}$ & $a$ (fm) & $m_\pi $ & $u_{0}$ & Phys. Vol. (fm$^4$)\\
\hline
$12^3\times{24}$ & 4.60  & 0.0000 & 0.120  & $\infty$ & 0.8888 & $1.44^3\times{2.88}$ \\
$16^3\times{32}$ & 4.80  & 0.0000 & 0.096  & $\infty$ & 0.8966 & $1.54^3\times{3.08}$ \\
$12^3\times{24}$ & 4.00  & 0.1318 & 0.120  & 806      & 0.8338 & $1.44^3\times{2.88}$ \\
$16^3\times{32}$ & 4.20  & 0.1300 & 0.096  & 820      & 0.8745 & $1.54^3\times{3.08}$ \\
\end{tabular}
\caption{Parameters for the different lattices. All use a tadpole improved Luscher Weisz gluon action. Lattice spacings are set via a string tension analysis incorporating the lattice Coulomb term. Shown is the lattice volume, gauge coupling $\beta,$ lattice spacing, pion mass (for the dynaical quarks) in MeV, the mean link and the physical volume.}
\label{tab:proplattices}
\end{ruledtabular}
\end{table}

\section{Simulation Details}

Calculations are performed on two quenched and two dynamical lattices. The dynamical lattices use a FLIC fermion action for the two degenerate flavors of sea quark. All the lattices have approximately the same physical volume, and the details of all lattices are given in Table \ref{tab:proplattices}. Two lattice volumes are used, $12^3\times 24$ and $16^3\times 32,$ with the lattice spacings for the quenched and dynamical lattices approximately matched. Landau gauge is chosen for the gauge fixing. An improved gauge fixing scheme~\cite{bowman2} is used, and a Conjugate Gradient Fourier Acceleration~\cite{cm} algorithm is chosen to perform the gauge fixing.

\begin{table}[b]
\centering
\begin{ruledtabular}
\begin{tabular}{ccccc}
$\mu$ & $a=0.120$ & $a=0.096$ \\
\hline
$\mu_1$ & 0.00400 & 0.00305 \\
$\mu_2$ & 0.00800 & 0.00610 \\
$\mu_3$ & 0.01200 & 0.00915 \\
$\mu_4$ & 0.01600 & 0.01220 \\
$\mu_5$ & 0.02000 & 0.01525 \\
$\mu_6$ & 0.02400 & 0.01830 \\
$\mu_7$ & 0.02800 & 0.02134 \\
$\mu_8$ & 0.03200 & 0.02439 \\
$\mu_9$ & 0.04000 & 0.03049 \\
$\mu_{10}$ & 0.04800 & 0.03659 \\
$\mu_{11}$ & 0.06000 & 0.04574 \\
$\mu_{12}$ & 0.08000 & 0.06098 \\
$\mu_{13}$ & 0.10000 & 0.07623 \\
$\mu_{14}$ & 0.12000 & 0.09148 \\
$\mu_{15}$ & 0.14000 & 0.10672 
\end{tabular}
\caption{The 15 lattice mass parameters $\mu$ used for each $a.$ Values are chosen so that the bare masses are approximately matched for each lattice.}
\label{tab:flicoverprop}
\end{ruledtabular}
\end{table}

Each lattice ensemble consists of 50 configurations. The lattice fermionic Green's function is obtained by inverting the FLIC overlap Dirac operator on each configuration using a multi-mass CG inverter~\cite{many-masses}.  We use periodic boundary conditions in the spatial directions and anti-periodic in the time direction. The Fourier transform is taken to convert to momentum space, and the the bare quark propagator is obtained from the ensemble average. The quark propagator is calculated for 15 different masses. The details of the FLIC overlap parameters used are presented in Table \ref{tab:flicoverprop}. The matrix sign function in the FLIC overlap is evaluated using the Zolotarev rational polynomial approximation~\cite{chiu-zolotarev}, of degree (typically 8) chosen to give an accuracy of $2.0\times10^{-8}$ within the spectral range of the kernel, after projecting out low-lying modes. The tree level propagator is calculated by setting the links and the mean link to unity (the free theory). The kinematical lattice momentum $q$ is obtained numerically from the tree level propagator, although it could equally well have been obtained from the analytic form for $q$ derived in Ref.~\cite{overlgp}.

The gluon propagator is obtained by performing a Fourier Transform of the lattice gluon field $A_\mu$ using the midpoint definition and then calculating the scalar propagator directly.

\section{Results}

\begin{figure*}[p]
\includegraphics[angle=90,height=0.28\textheight,width=0.45\textwidth]{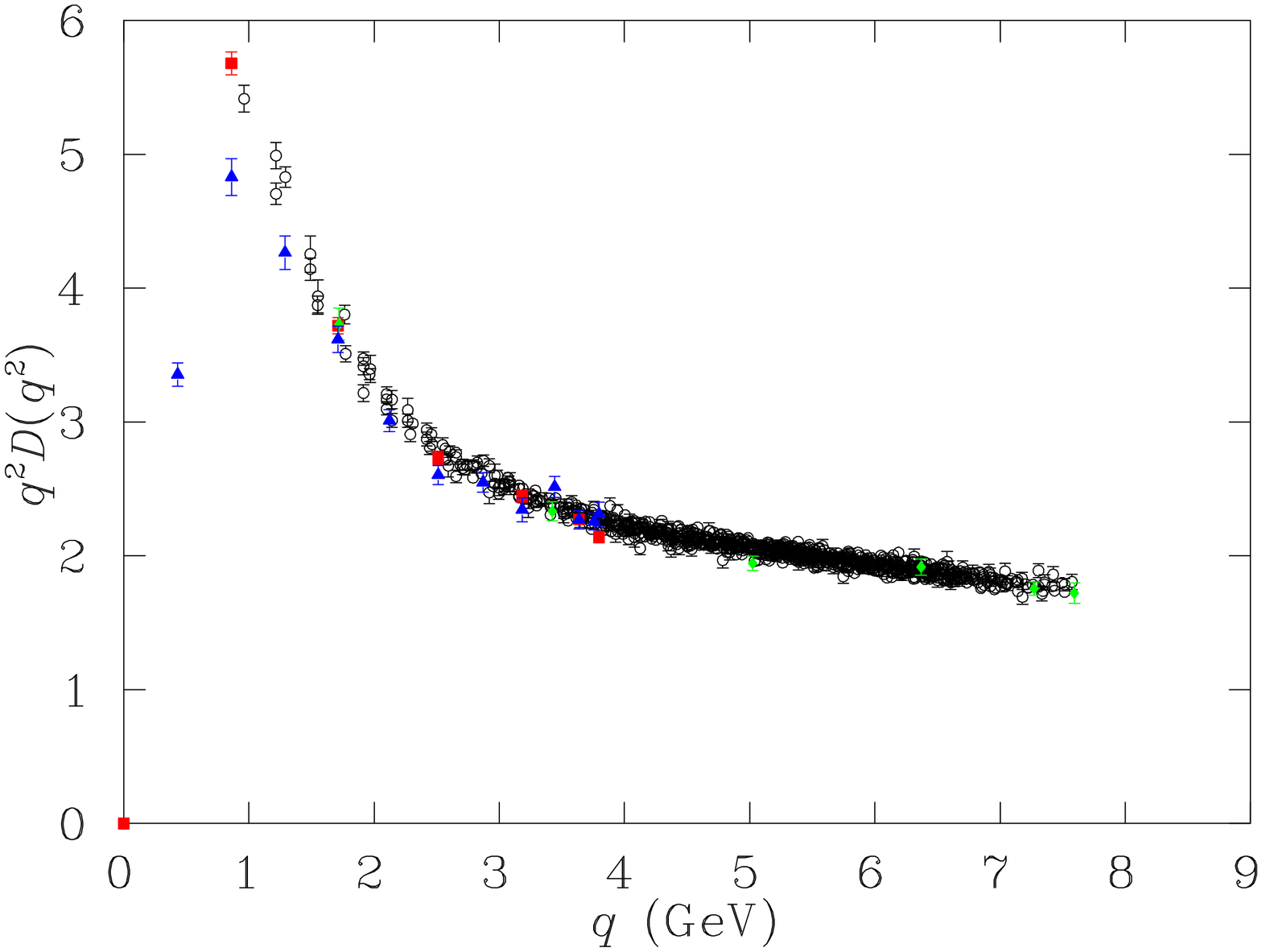}
\includegraphics[angle=90,height=0.28\textheight,width=0.45\textwidth]{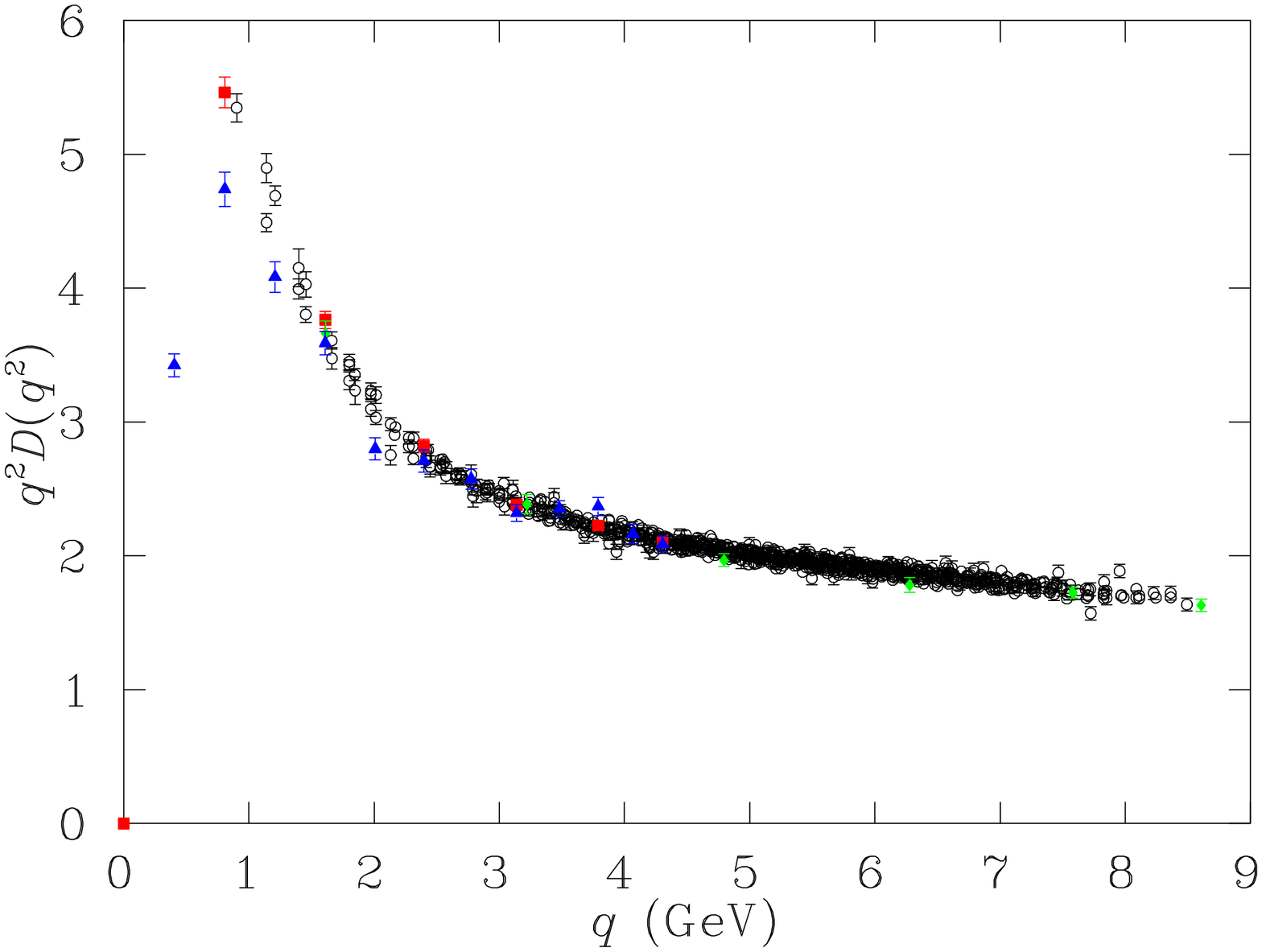}

\includegraphics[angle=90,height=0.28\textheight,width=0.45\textwidth]{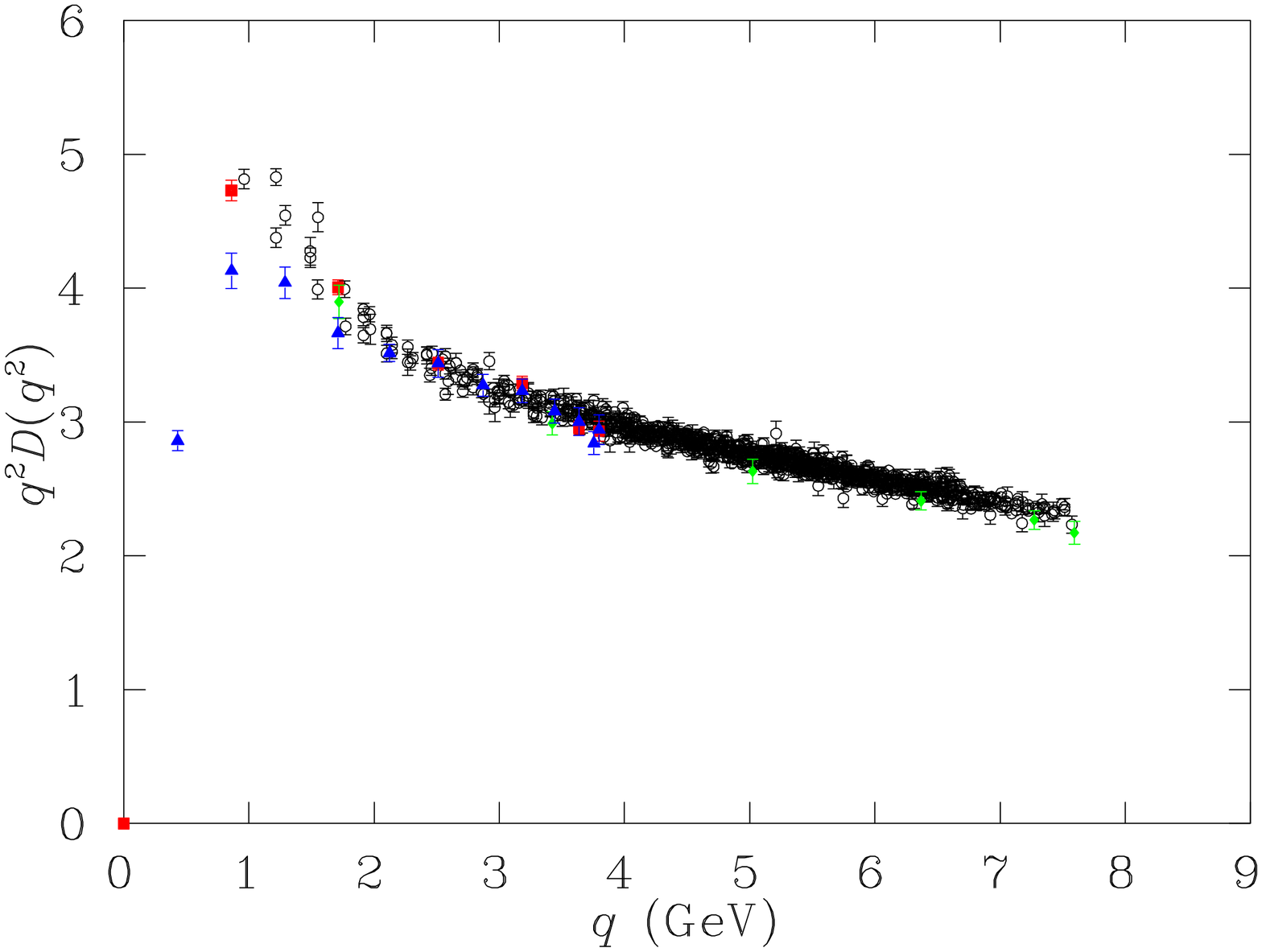}
\includegraphics[angle=90,height=0.28\textheight,width=0.45\textwidth]{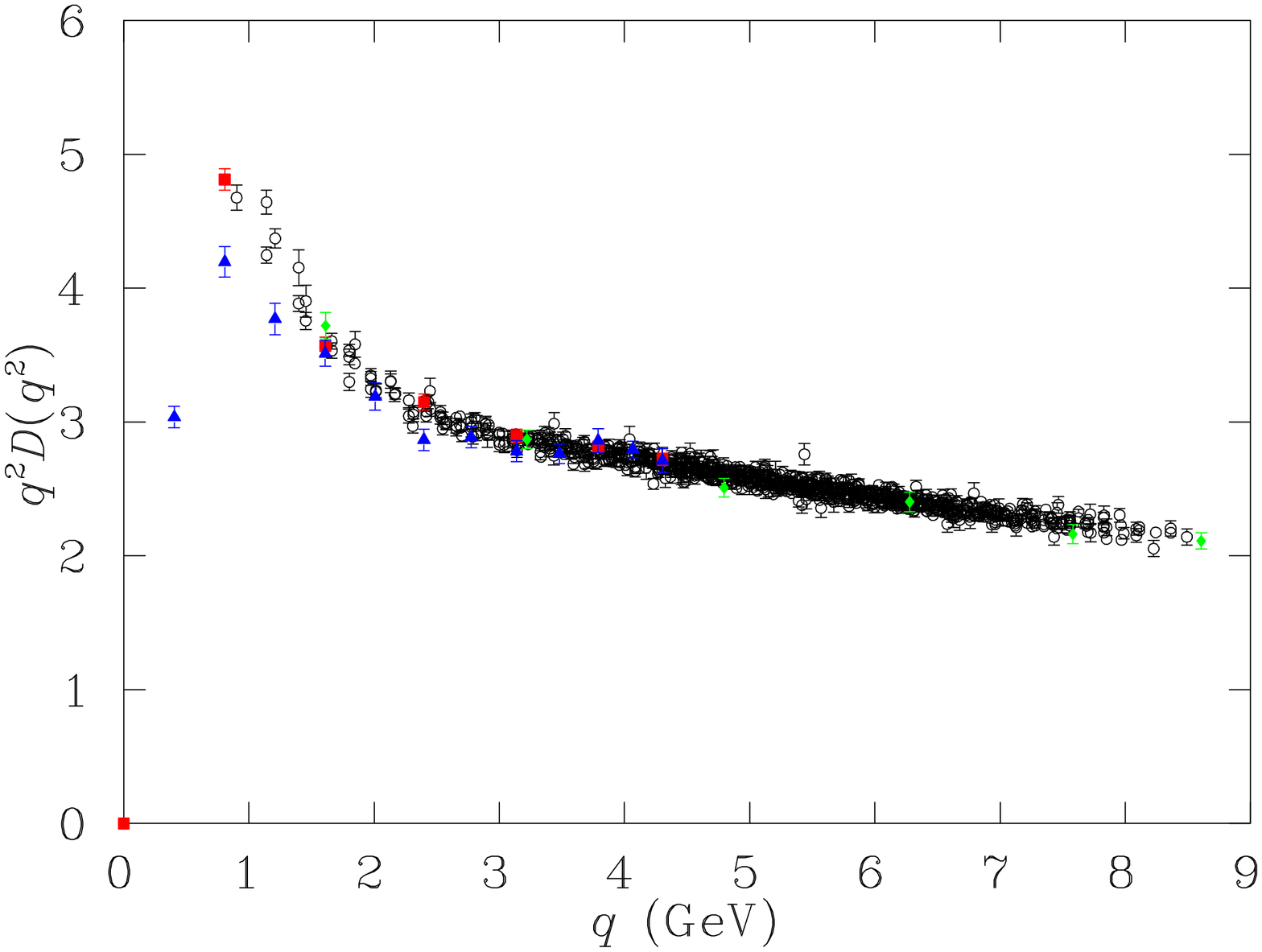}

\caption{Gluon propagator data for the bare gluon dressing function $q^2 D(q^2).$ The quenched lattices are shown on the top and the dynamical lattices on the bottom. The left column shows the results for $a=0.120$ and the right column for $a=0.096.$}
\label{fig:gluondata}
\end{figure*}

\begin{figure*}[p]

\includegraphics[angle=90,height=0.28\textheight,width=0.45\textwidth]{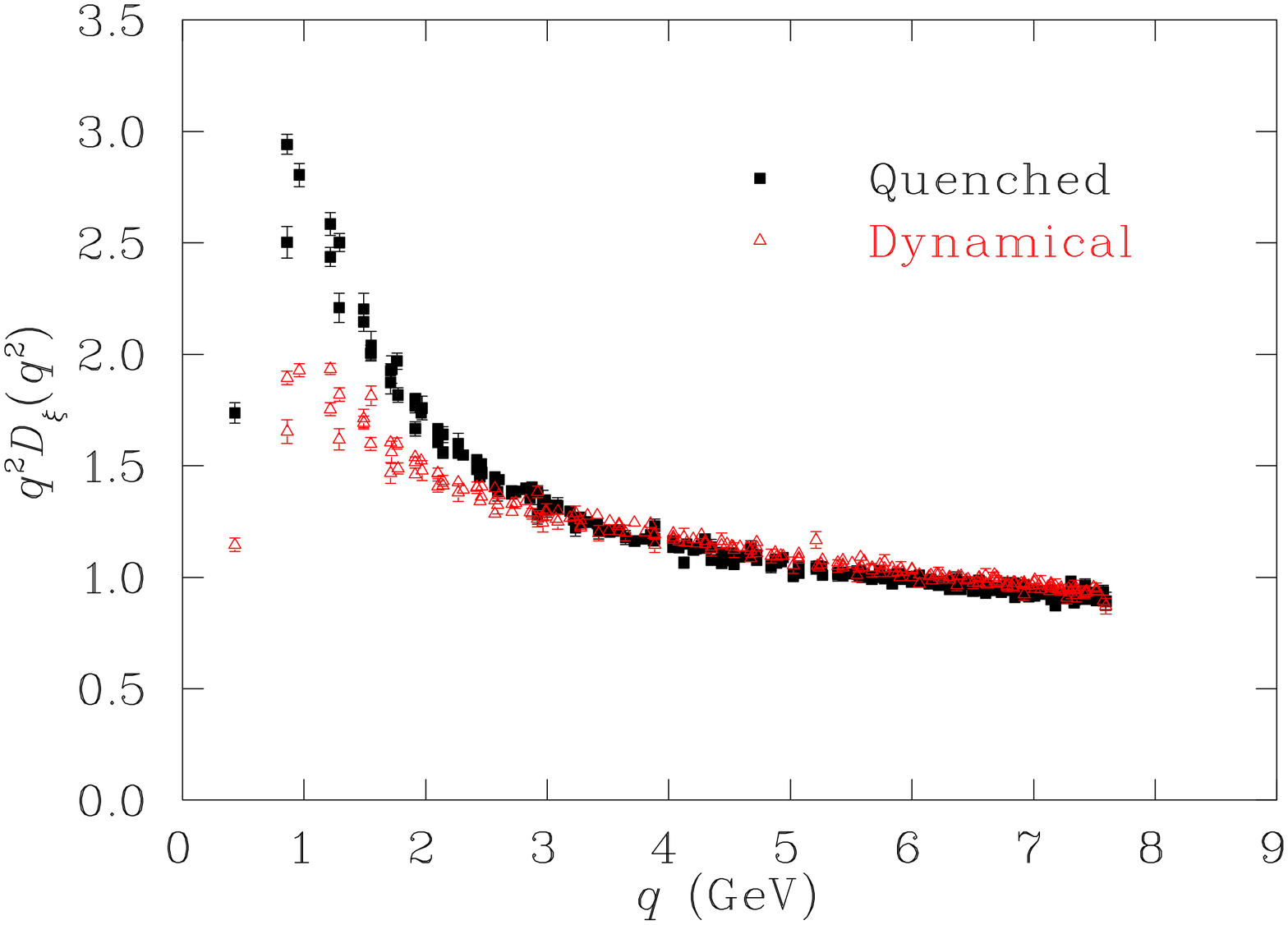}
\includegraphics[angle=90,height=0.28\textheight,width=0.45\textwidth]{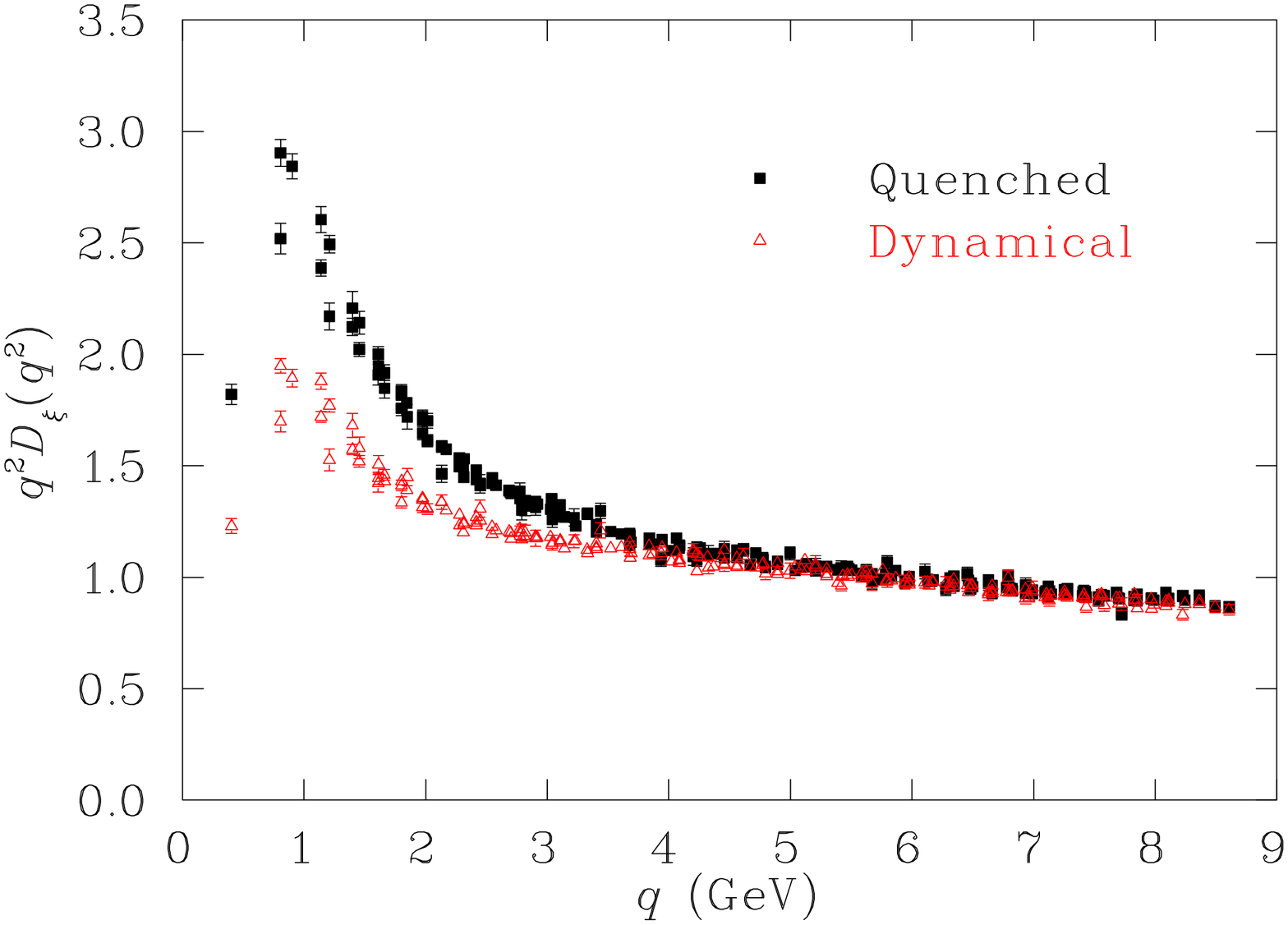}
\caption{Plots comparing the quenched (squares) and dynamical (triangles) renormalised gluon dressing function, $q^2D_\xi(q^2),$ for $a=0.120$ (left) and $a=0.096$ (right). The renormalization point $\xi$ was chosen to be 6.0 GeV. }
\label{fig:gluoncomp}
\end{figure*}

We first examine the gluon propagator results. Figure~\ref{fig:gluondata} shows
the bare gluon dressing function $q^2 D(q^2)$ on each of the lattices. The 
squares represent momenta entirely in a spatial cartesian direction while
triangles indicate points where the momentum is entirely along the temporal 
axis.  The separation between these is due to the use of an asymmetric lattice
and is a clear indication of finite volume effects.  This is to be expected on
such small physical volumes. 
The quenched and dynamical renormalised gluon dressing functions 
$q^2D_\xi(q^2)$ are compared directly in Figure~\ref{fig:gluoncomp}. We 
see that the presence of dynamical quarks causes the suppression of the 
dressing function in the infrared.  This effect is clear even with the 
relatively heavy sea quarks used in this study.  The same effect was seen in 
work which 
compared the gluon propagator on quenched and dynamical lattices with a 
staggered fermion action\cite{unquenched-gluon,unquenched-quark}.

\begin{figure*}[p]
\includegraphics[angle=90,height=0.28\textheight,width=0.45\textwidth]{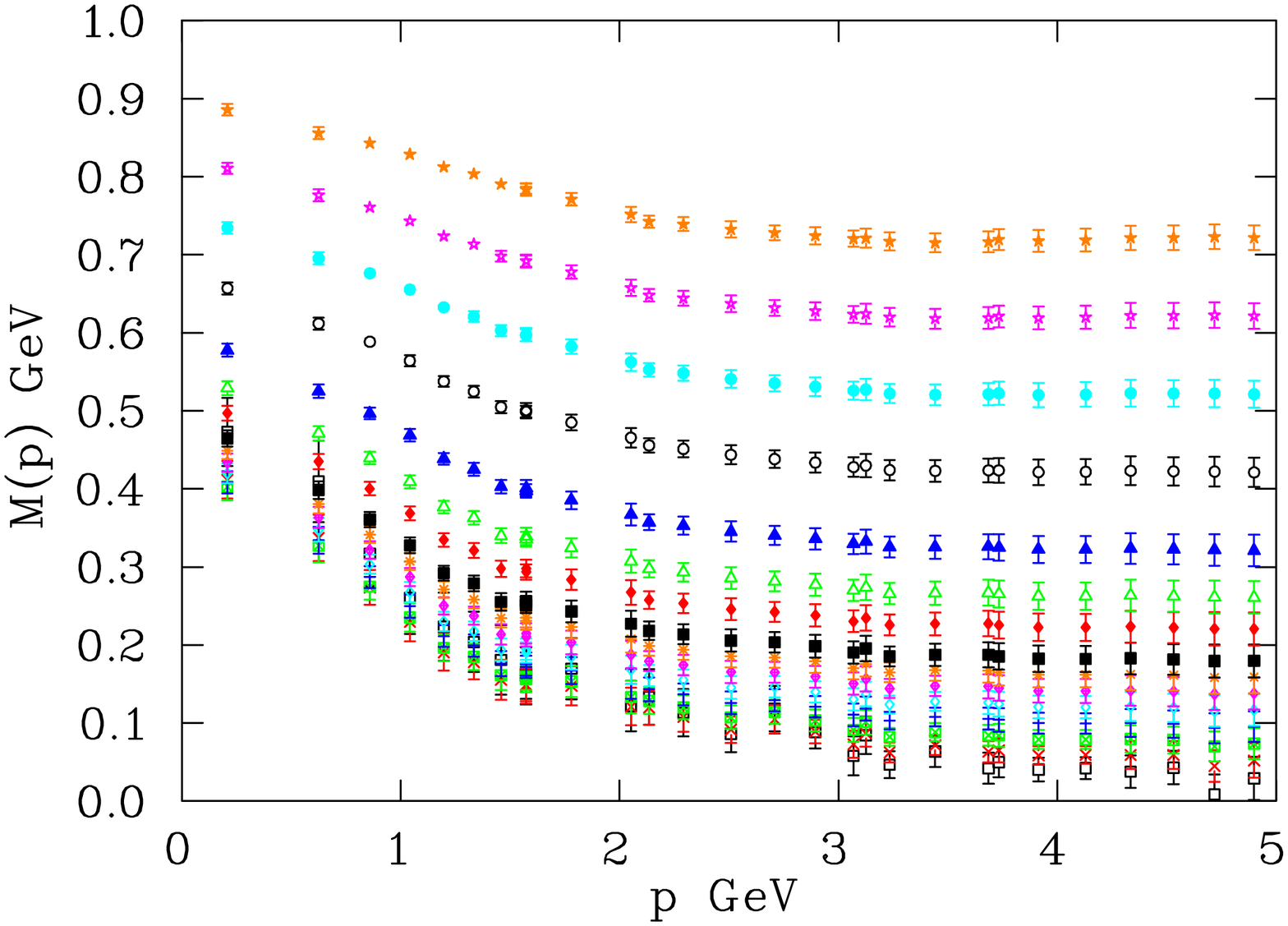}
\includegraphics[angle=90,height=0.28\textheight,width=0.45\textwidth]{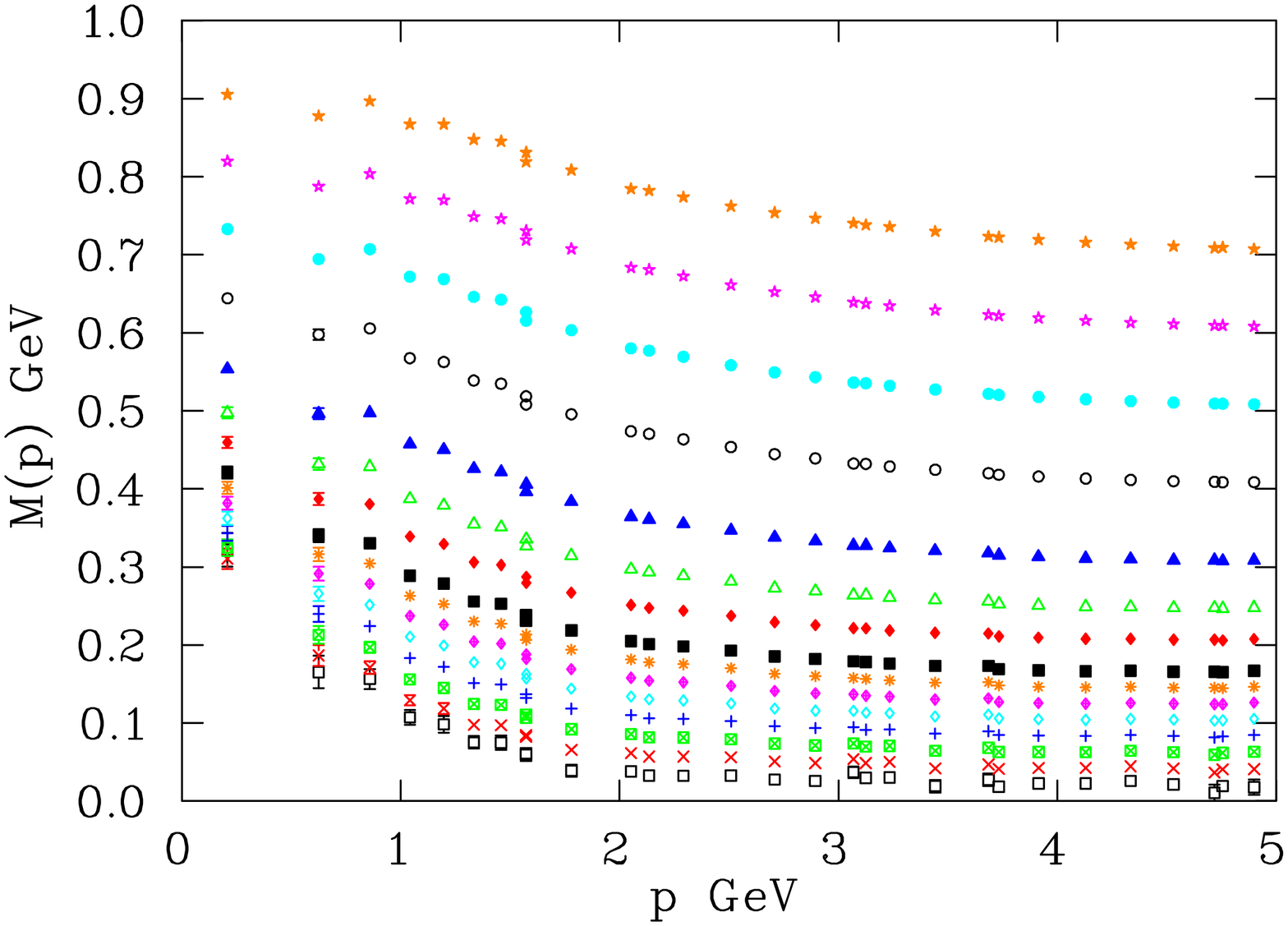}

\caption{Cylinder cut data for the dynamical FLIC Overlap mass function $M(p)$ at finite quark mass for the for the lattices at $a=0.120$ (left) and $a=0.096$ (right). The plots are against the discrete lattice momentum $p.$}
\label{fig:massfunc}
\end{figure*}

\begin{figure*}[p]
\includegraphics[angle=90,height=0.28\textheight,width=0.45\textwidth]{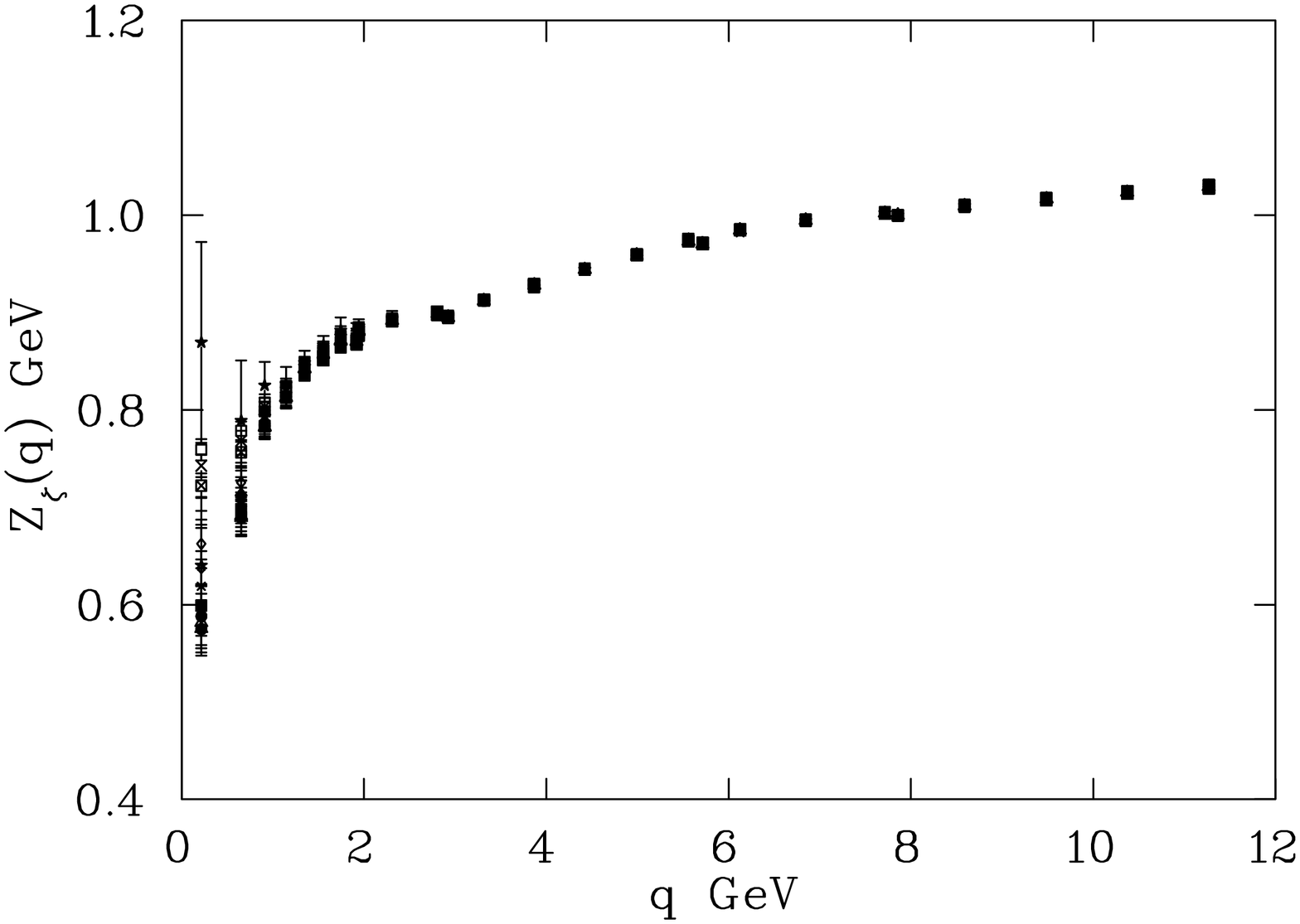}
\includegraphics[angle=90,height=0.28\textheight,width=0.45\textwidth]{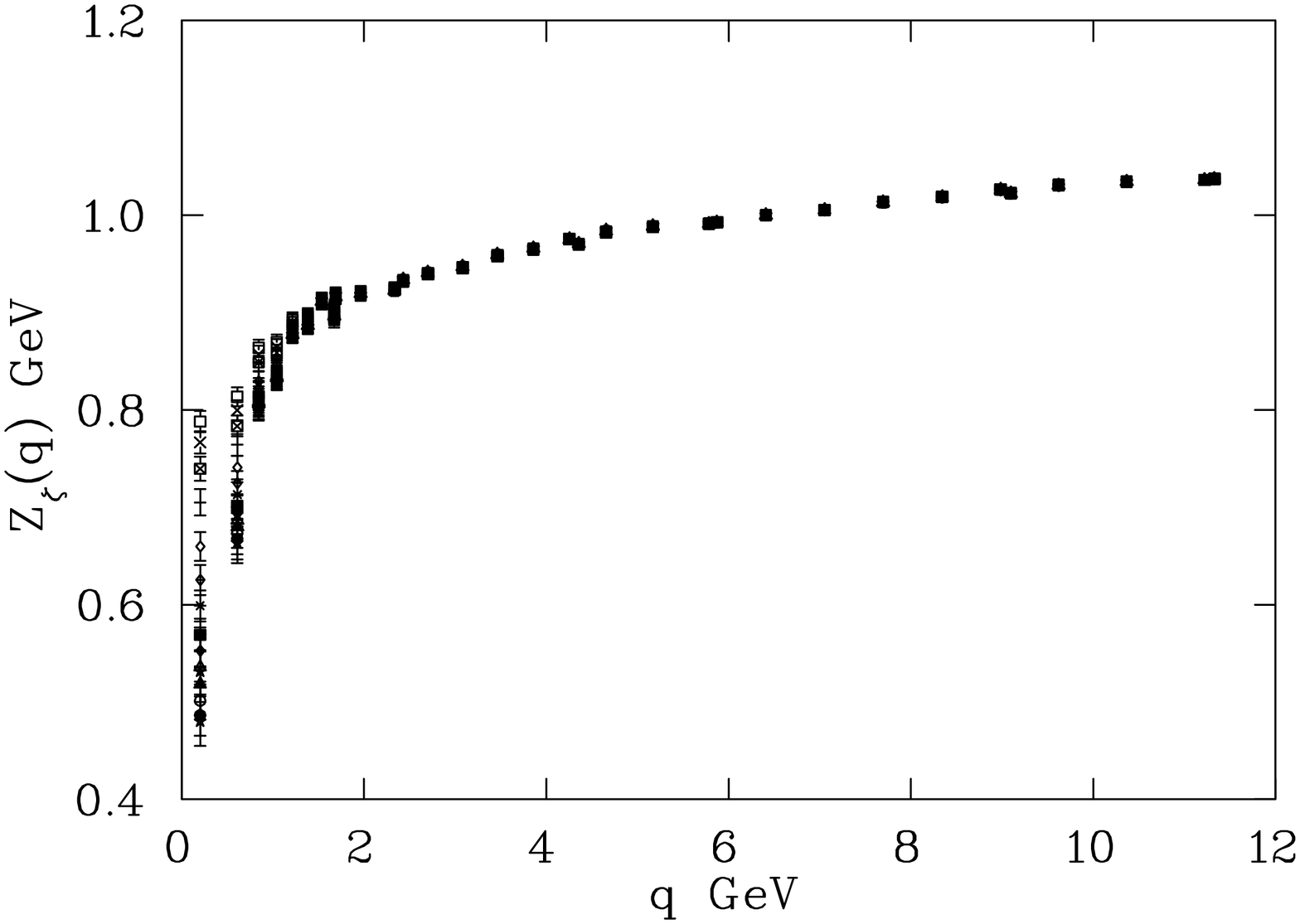}

\caption{Cylinder cut data for the dynamical renormalization function $Z_\zeta(q)\ (\zeta=6~\text{GeV})$ at finite quark mass for the lattices at $a=0.120$ (left) and $a=0.096$ (right). The plots are against the kinematical lattice momentum $q.$} 
\label{fig:zfunc}
\end{figure*}

We now turn to the quark propagator results. The mass function $M(p)$ and the renormalization function $Z_\zeta(p)$ are calculated on each of the lattices. A cylinder cut~\cite{overlgp} is applied to all the data, to reduce the effects of rotational symmetry violation. The full results for the dynamical lattices are displayed in Fig.~\ref{fig:massfunc} for $M(p)$ and Fig.~\ref{fig:zfunc} for $Z_\zeta(p).$  The renormalization point $\zeta$ for $Z_\zeta(p)$ is chosen to be approximately $q=6\text{ GeV.}$ The full results for the quenched FLIC overlap quark propagator presented in previous work\cite{kamleh-overlap}.

\begin{figure*}[p]
\centering
\includegraphics[angle=90,height=0.28\textheight,width=0.45\textwidth]{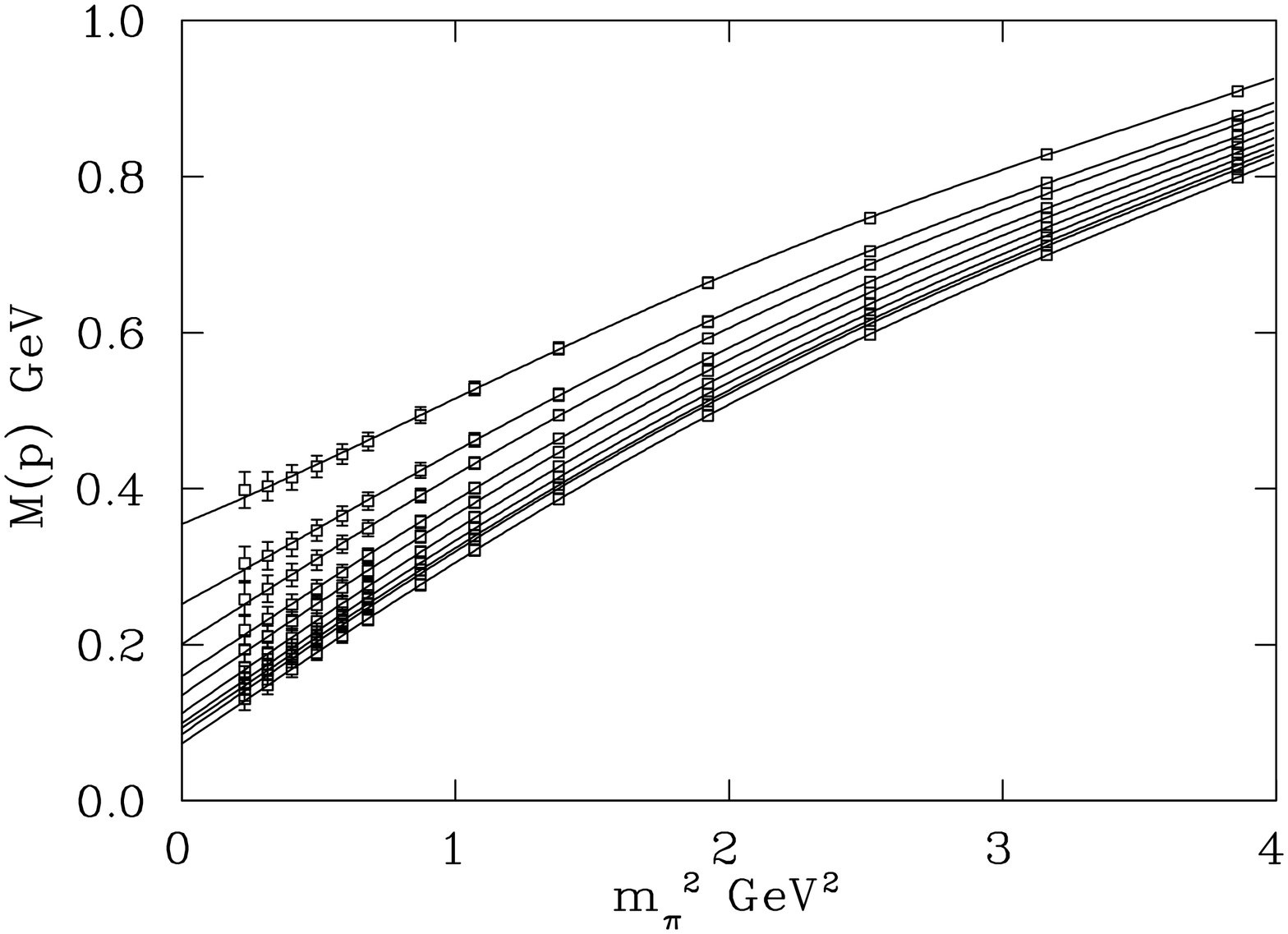}
\includegraphics[angle=90,height=0.28\textheight,width=0.45\textwidth]{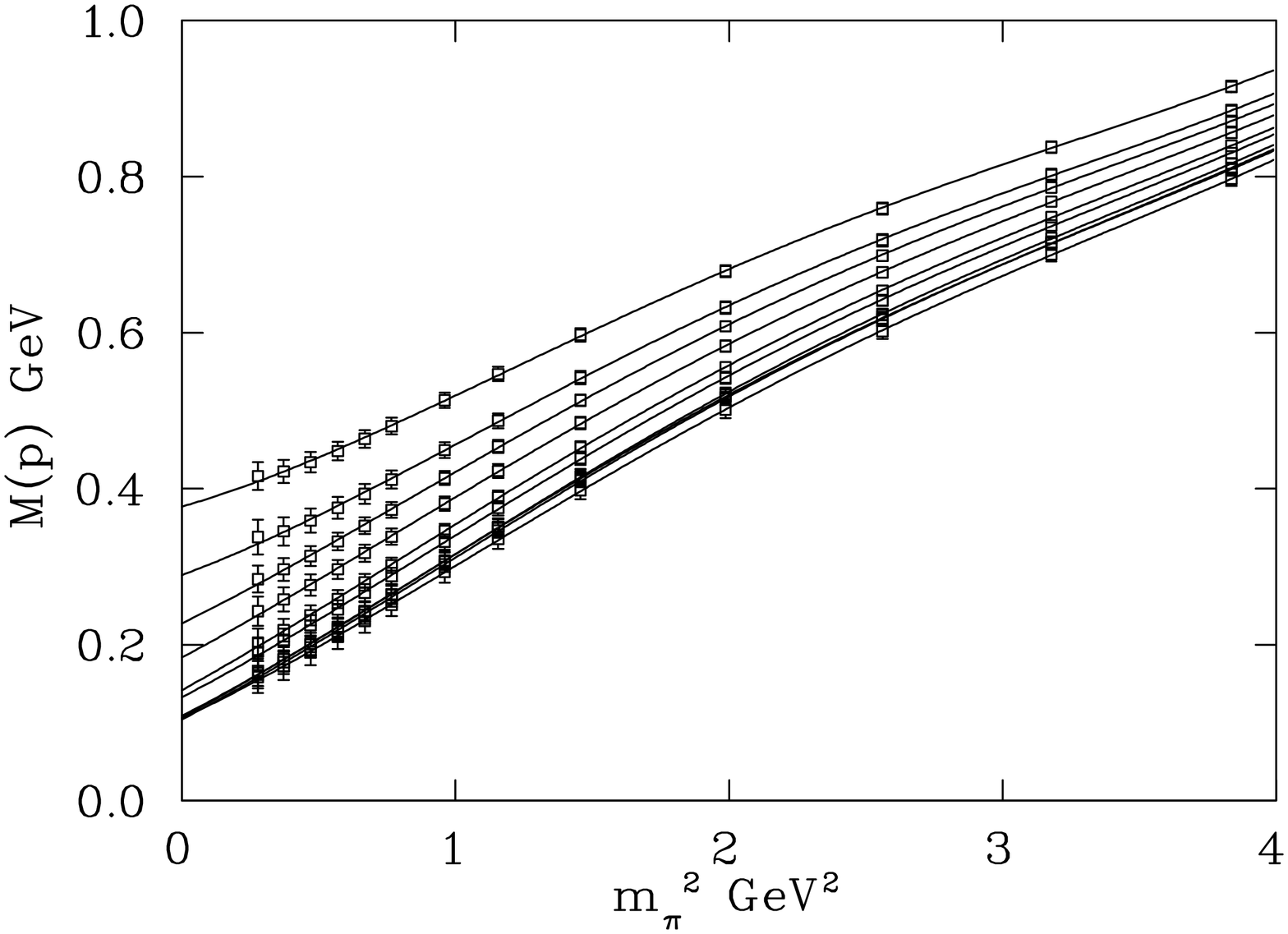}

\includegraphics[angle=90,height=0.28\textheight,width=0.45\textwidth]{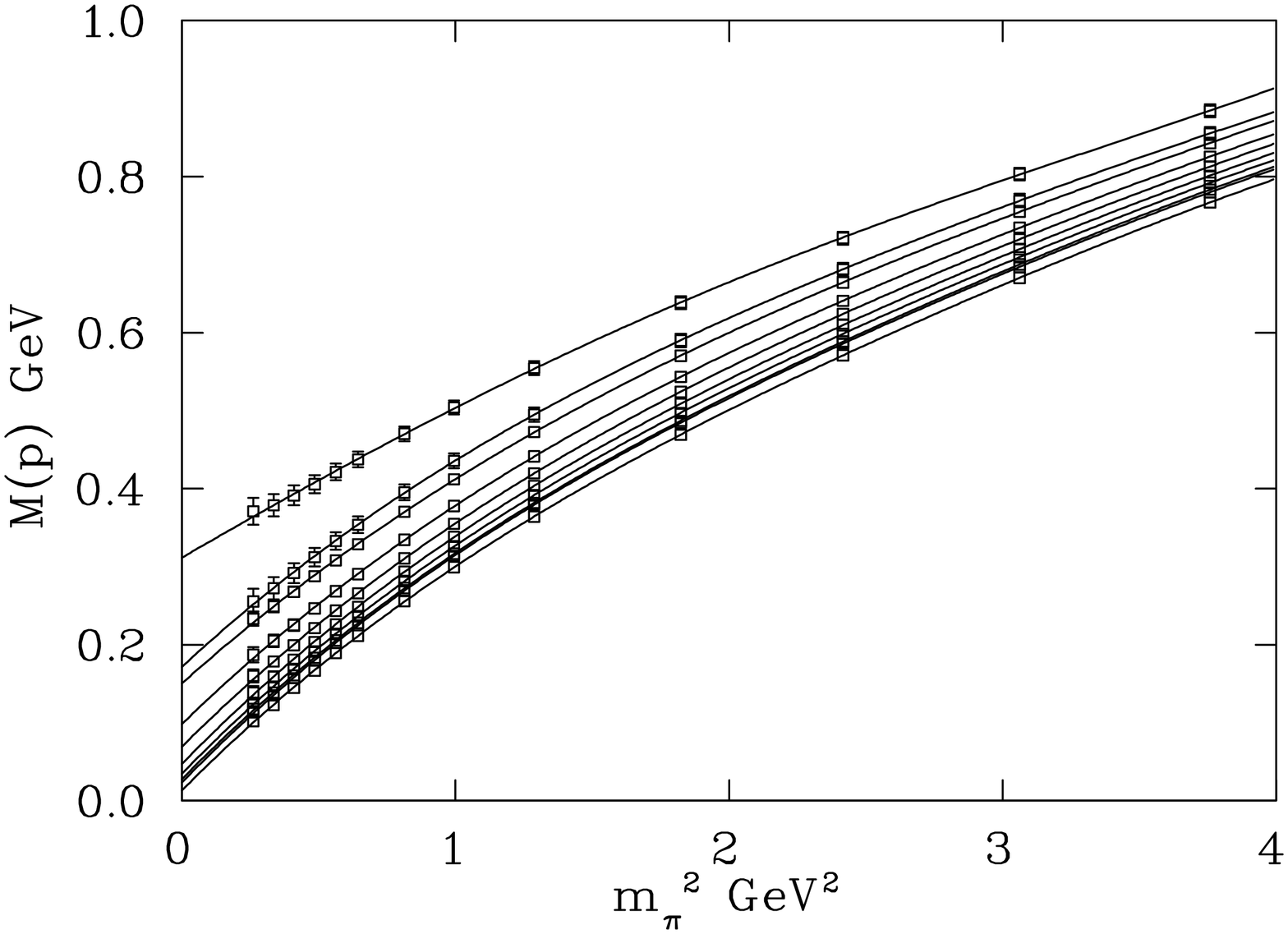}
\includegraphics[angle=90,height=0.28\textheight,width=0.45\textwidth]{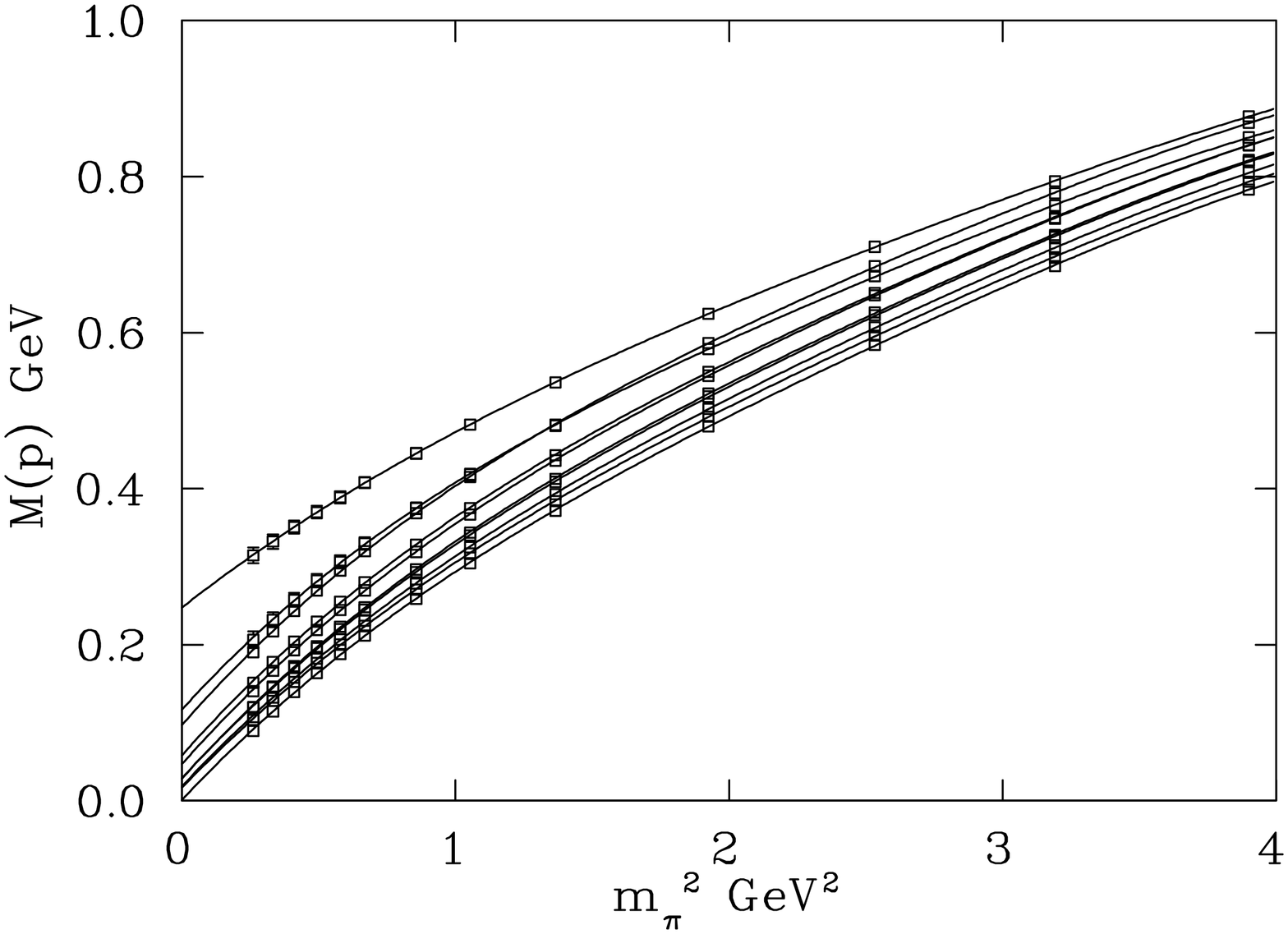}

\caption{Cylinder cut data showing the dependence of the mass function $M(p)$ for the quenched (left) and dynamical (right) FLIC Overlap on the bare mass $m^0$ at fixed momenta, for the lowest 10 momenta values. At small bare mass, the curves are ordered inversely to the momenta they represent, that is the smallest momenta is the topmost curve. Data is shown for the lattices at $a=0.120$ (top) and $a=0.096$ (bottom). The solid curves are the quartic fits to the data.}
\label{fig:mqchifit}
\end{figure*}

\begin{figure*}[p]
\centering
\includegraphics[angle=90,height=0.28\textheight,width=0.45\textwidth]{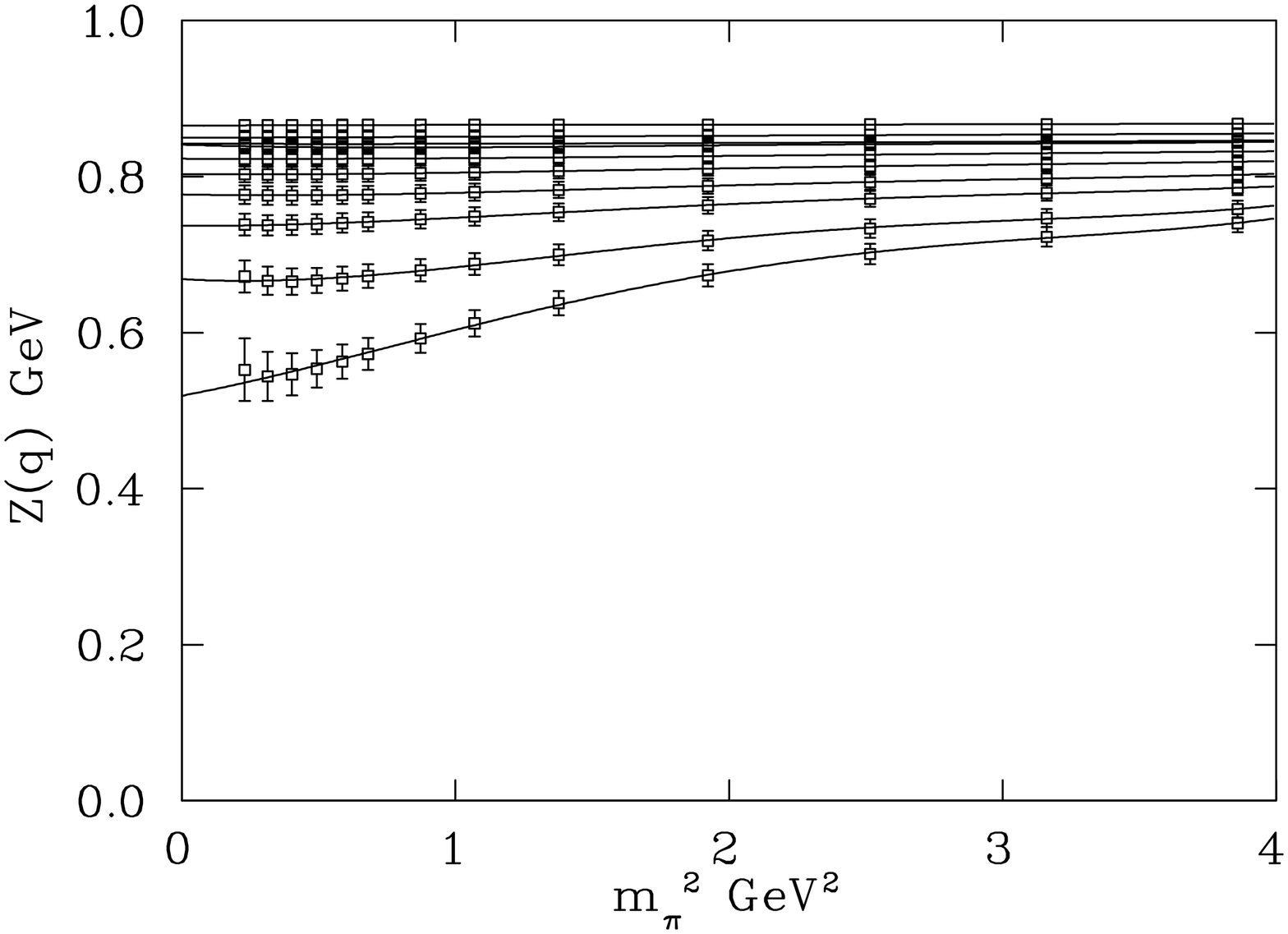}
\includegraphics[angle=90,height=0.28\textheight,width=0.45\textwidth]{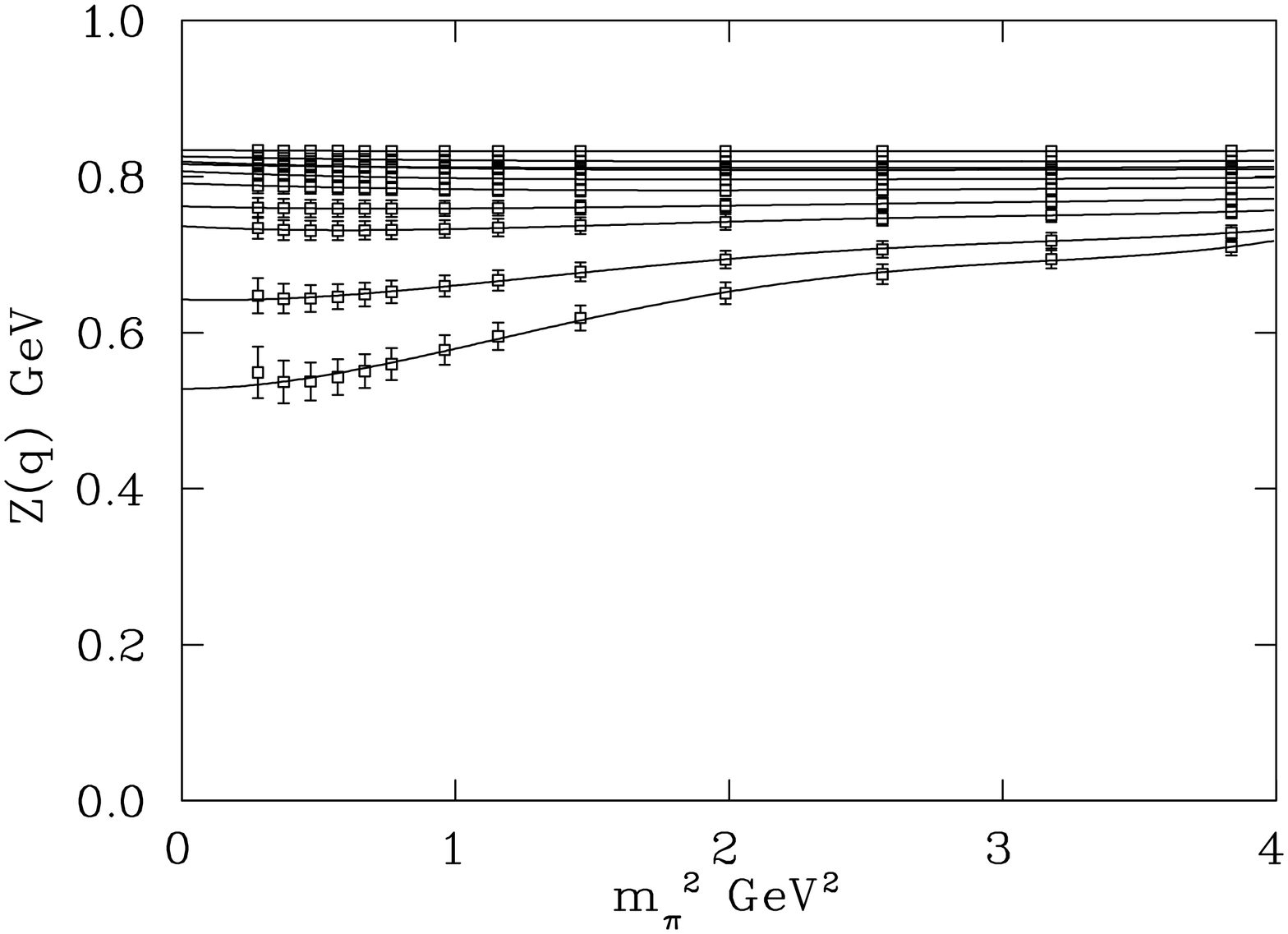}

\includegraphics[angle=90,height=0.28\textheight,width=0.45\textwidth]{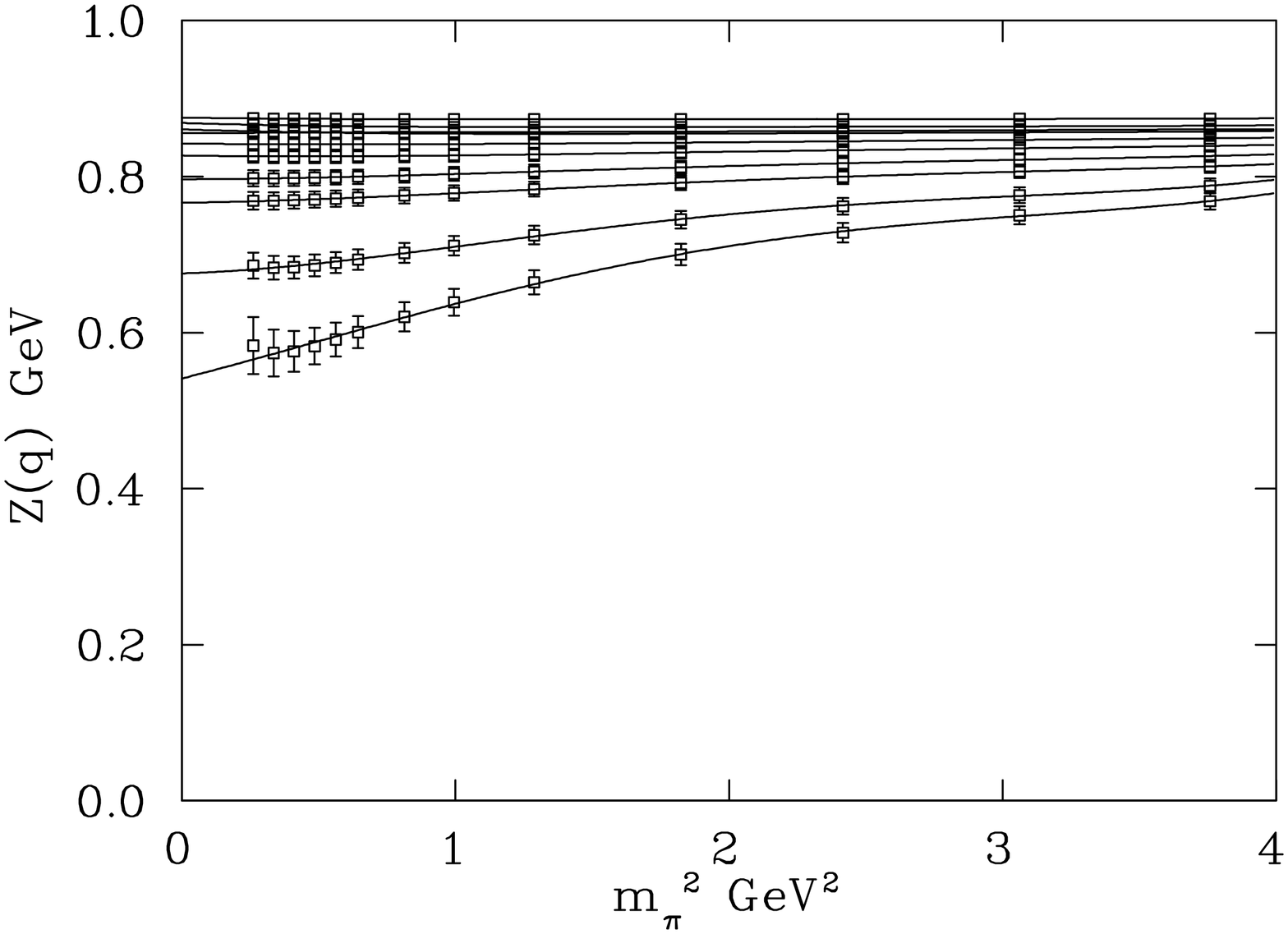}
\includegraphics[angle=90,height=0.28\textheight,width=0.45\textwidth]{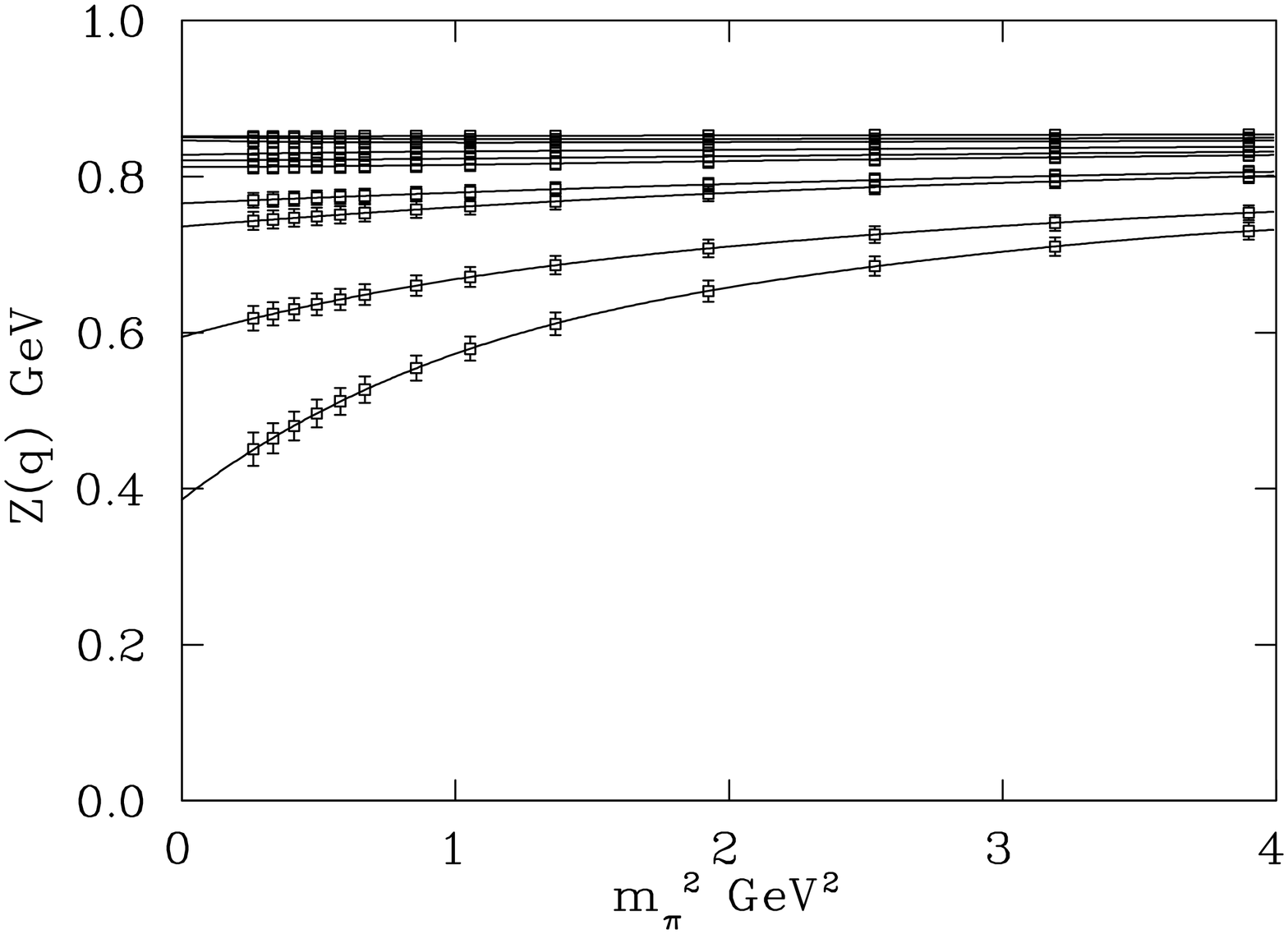}

\caption{Cylinder cut data showing the dependence of the renormalization function $Z(p)$ for the quenched (left) and dynamical (right) FLIC Overlap on the pion mass $m_\pi^2$ at fixed momenta, for the lowest 10 momenta values. The curves are ordered according to the momenta they represent, that is the largest momenta is the topmost curve. Data is shown for for the lattices at $a=0.120$ (top) and $a=0.096$ (bottom). The solid curves are the quartic fits to the data.}
\label{fig:zqchifit}
\end{figure*}

\begin{figure*}[p]
\centering
\includegraphics[angle=90,height=0.28\textheight,width=0.45\textwidth]{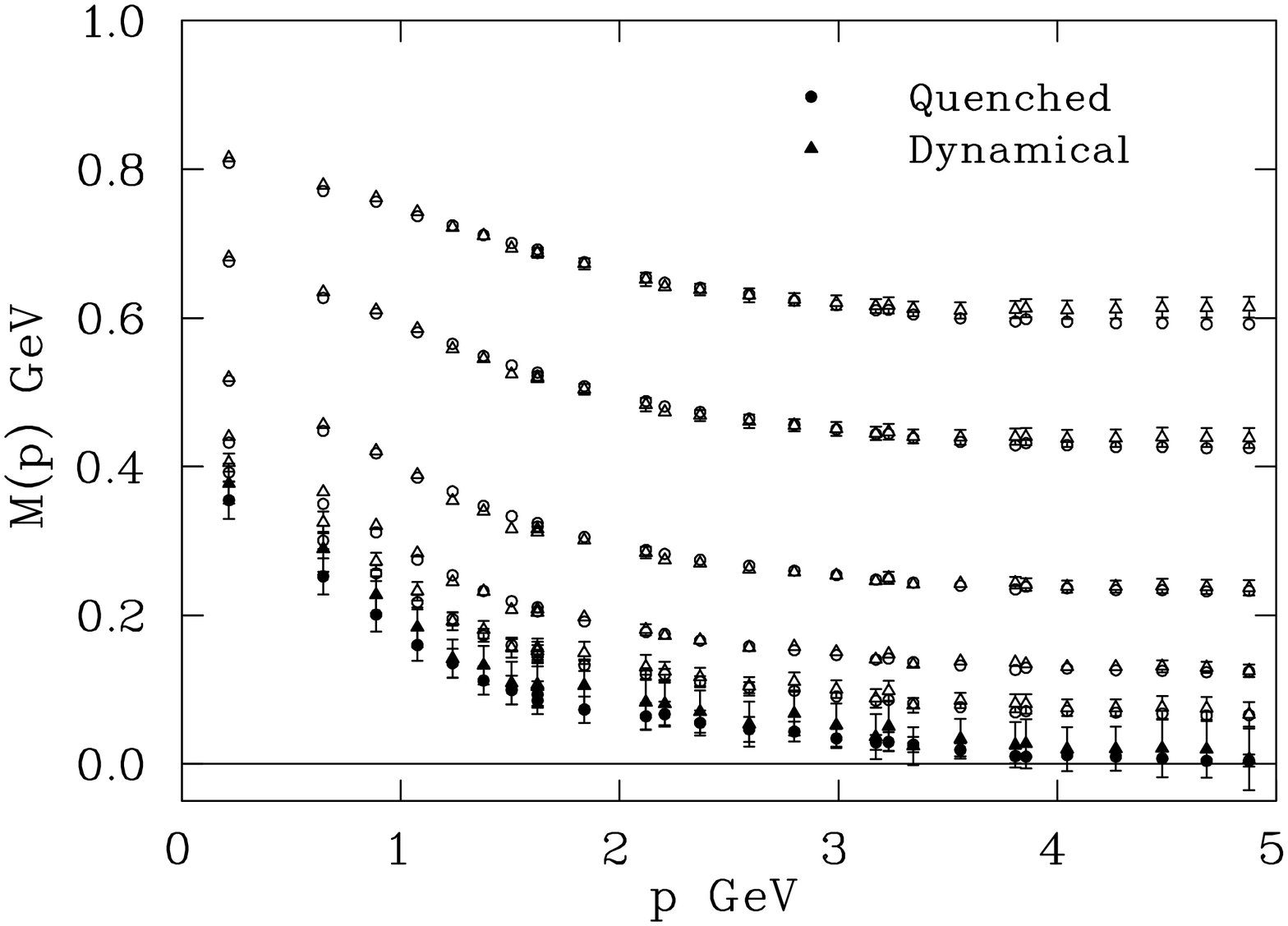}
\includegraphics[angle=90,height=0.28\textheight,width=0.45\textwidth]{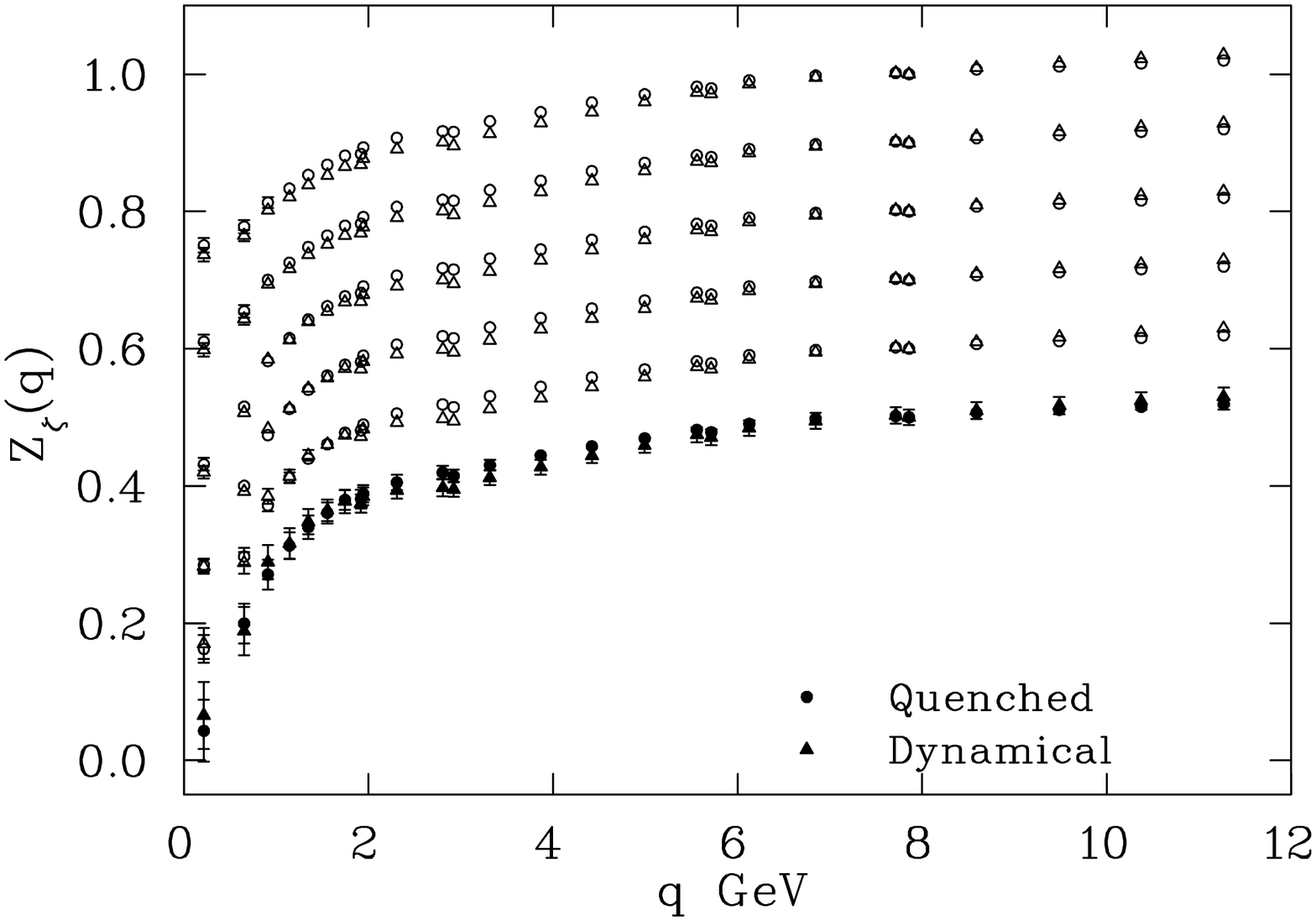}

\includegraphics[angle=90,height=0.28\textheight,width=0.45\textwidth]{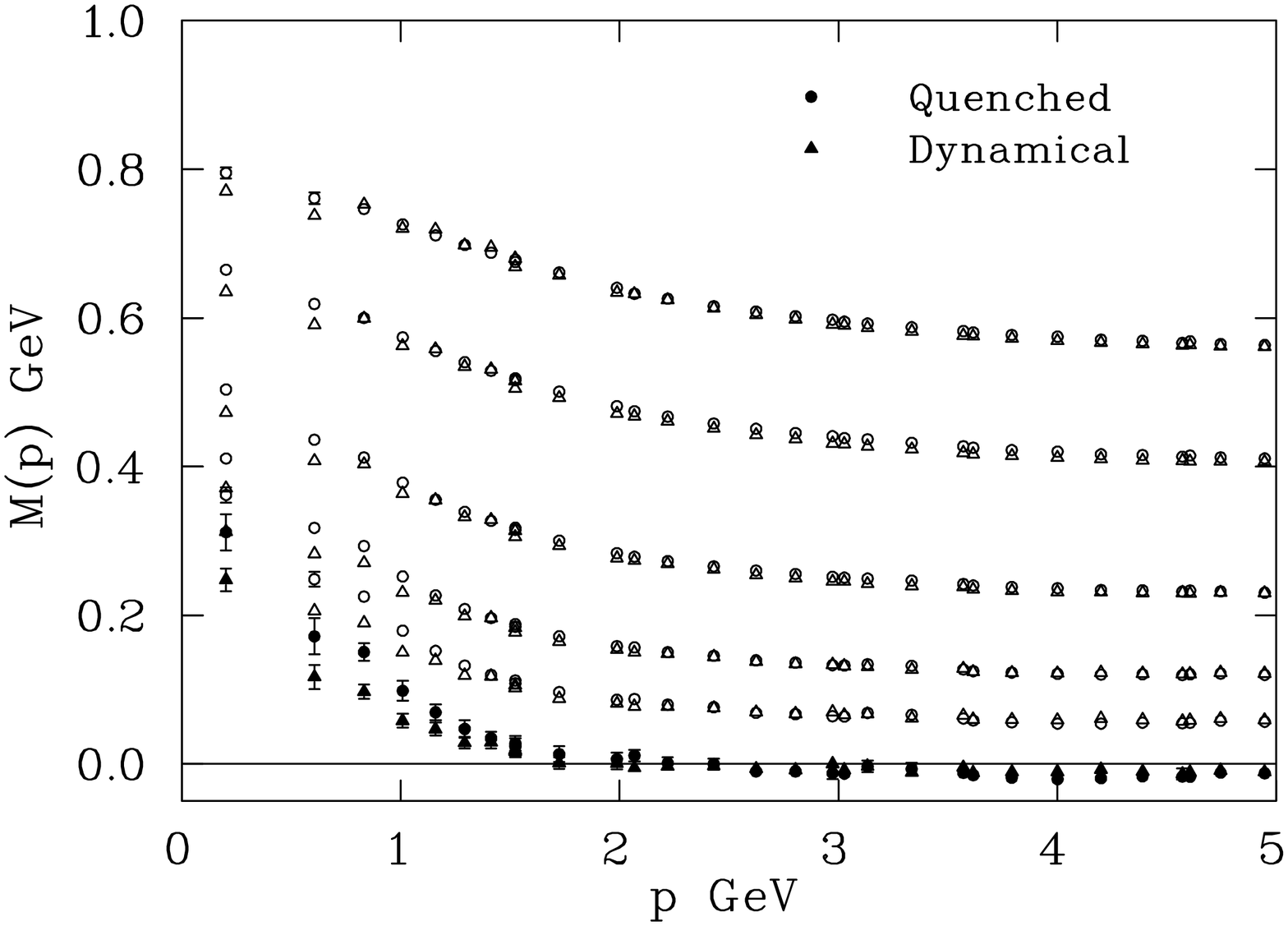}
\includegraphics[angle=90,height=0.28\textheight,width=0.45\textwidth]{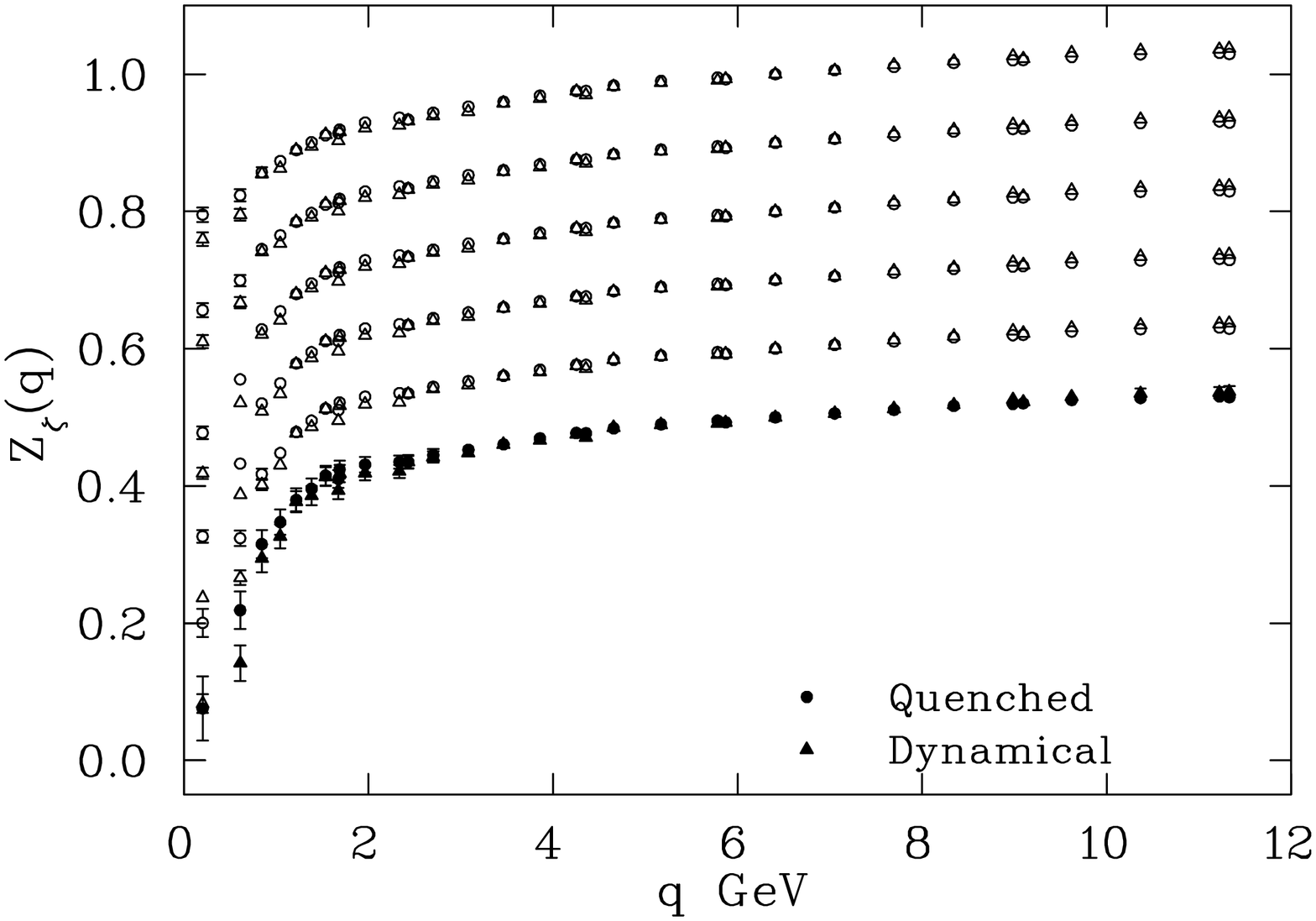}

\caption{Cylinder cut data comparing the interpolated mass function $M(p)$ (left) and renormalization function $Z_\zeta(q)$ (right), for the dynamical and quenched lattices at $a=0.120$ (top), $a=0.096$ (bottom). The different curves from lowest to highest represent matched pion masses of $m_\pi^2 = 0.0, 0.25, 0.5, 1.0, 2.0, 3.0\text{ GeV}^2.$ The solid points indicate the chiral limit. The different $Z_\zeta(q)$ have been offset vertically for clarity. The renormalization point is chosen to be $\zeta\approx 6\text{ GeV.}$ }
\label{fig:flicvsflic}
\end{figure*}

In order to compare the quenched and dynamical results, for each fixed momenta value we perform a quartic fit of both $M(p)$ and $Z(p)$ as a function of the pion mass (excluding the lightest two masses for finite volume reasons). The data and fits for the smallest 10 momenta on each of the lattices are shown in Fig.~\ref{fig:mqchifit} for $M(p)$ and Fig.~\ref{fig:zqchifit} for $Z(p).$ Having obtained the fit functions $M(p,m_\pi^2)$ and $Z(p,m_\pi^2)$ we then choose $m_\pi^2 = 0.0, 0.25, 0.5, 1.0, 2.0, 3.0\text{ GeV}^2$ and compare the functions at these values of the pion mass. In this way we can compare the quenched and dynamical results with matched pion masses. Results are shown in Figure \ref{fig:flicvsflic}, to which we now turn our attention.

First we examine the mass function. The results for $a=0.120$ show little effect due to the presence of sea quarks, just a slight suppression in the infrared for the lighter masses. The heavier masses show essentially no difference between the quenched and dynamical results, execpt in the extreme ultraviolet, where the discretisation errors give some difference. The discretisation errors are not present in the $a=0.096$ results, but now we can clearly see infrared suppression of $M(p)$ in the dynamical results for all masses. 

Turning to the renormalization function, both the lattices show infrared suppresssion of $Z_\zeta(q)$ in the presence of dynamical quarks, and agreement in the far ultraviolet. At $a=0.096$ the quenched and dynamical results also agree in the intermediate momentum regime, whereas there is a slight discrepancy in this region at $a=0.120.$

\section{Conclusions}

The gluon propagator and FLIC overlap quark propagator are compared on quenched and dynamical FLIC lattices at $a=0.096$ and $a=0.120.$ We observe the the presence of dynamical quarks causes suppression in the infrared in all quantities examined: the gluon dressing function $q^2 D(q^2),$ the mass function $M(p)$ and the quark renormalization function $Z(q).$ Some differences between the results for the two lattice spacings indicate we are not quite in the scaling region at the coarser spacing. An alternative source of systematic error in the the quark propagator results is the simplistic fitting functions used, which do not take into account the possibility of non-analytic behaviour at light quark mass. However, our results are consistent with those obtained elsewhere\cite{christian-qprop,unquenched-gluon,unquenched-quark}.

\acknowledgments

We thank both the South Australian Partnership for Advanced Computing (SAPAC)
and the Australian Partnership for Advanced Computing (APAC)
for generous grants of supercomputer time which have enabled this
project.  This work is supported by the Australian Research Council. WK is supported by SFI basic research grant 04/BR/P0266. JBZ is partly supported by Chinese NSFC-Grant No. 10675101.


\end{document}